\documentclass[11pt,a4paper]{article}
\pdfoutput=1 %%need when submit to JHEP
\usepackage{jheppub}
\usepackage{tikz}
\DeclareMathOperator{\Tr}{Tr}

\DeclareMathOperator{\diag}{diag}

\newcommand{\ri}{\mathrm{i}}

\newcommand{\cob}{\delta}

\newcommand{\hf}{\frac{1}{2}}
\newcommand{\qu}{\frac{1}{4}}
\newcommand{\til}[1]{\widetilde{#1}}

\renewcommand{\b}[1]{\overline{#1}}
\newcommand{\del}{\partial}

\newcommand{\bra}{\langle}
\newcommand{\ket}{\rangle}
\newcommand{\la}{\lambda}
\newcommand{\La}{\Lambda}
\newcommand{\ka}{\kappa}
\newcommand{\h}[1]{\widehat{#1}}
\newcommand{\bt}{\beta}
\newcommand{\ga}{\gamma}
\newcommand{\Ga}{\Gamma}

\newcommand{\rt}[1]{\sqrt{#1}}

\newcommand{\bbZ}{{\mathbb Z}}

\newcommand{\cF}{{\cal F}}
\newcommand{\cR}{{\cal R}}

\newcommand{\JT}{{\mbox{\scriptsize JT}}}

\newcommand{\Sch}{{\mbox{\scriptsize Sch}}}
\newcommand{\disk}{{\mbox{\scriptsize disk}}}
\newcommand{\trumpet}{{\mbox{\scriptsize trumpet}}}
\newcommand{\gs}{g_{\mbox{\scriptsize s}}}
\newcommand{\Dpartial}{D}
\newcommand{\tpartial}{\tilde{\partial}}
\newcommand{\tz}{\tilde{z}}

\begin{document}

\title{JT gravity, KdV equations and macroscopic loop operators}

\author[a]{Kazumi Okuyama}
\author[b]{and Kazuhiro Sakai}

\affiliation[a]{Department of Physics, Shinshu University,\\
3-1-1 Asahi, Matsumoto 390-8621, Japan}
\affiliation[b]{Institute of Physics, Meiji Gakuin University,\\
1518 Kamikurata-cho, Totsuka-ku, Yokohama 244-8539, Japan}

\emailAdd{kazumi@azusa.shinshu-u.ac.jp, kzhrsakai@gmail.com}

\abstract{
We study the thermal partition function of Jackiw-Teitelboim (JT)
gravity in asymptotically Euclidean $AdS_2$ background
using the matrix model description recently found
by Saad, Shenker and Stanford [arXiv:1903.11115].
We show that the partition function of JT gravity is written as the
expectation value of a macroscopic loop operator in the old matrix model
of 2d gravity in the background where infinitely many couplings
are turned on in a specific way. Based on this expression
we develop a very efficient method of computing the partition function
in the genus expansion as well as in the low temperature expansion
by making use of the Korteweg-de Vries constraints obeyed by
the partition function. We have computed both these expansions up to
very high orders using this method. It turns out that we can take a low
temperature limit with the ratio of the temperature and the genus
counting parameter held fixed. We find the first few orders of the
expansion of the free energy in a closed form in this scaling limit.
We also study numerically the behavior of the eigenvalue density and
the Baker-Akhiezer function using the results in the scaling limit.
}

\maketitle

%%%%%%%%%%%%%%%%%%%%%%%%%%%%%%%%%%%%%%%%%%%%%%%%%%%%%%%%%%%%%%%%%%%%%%%%
\section{Introduction \label{sec:intro}}
%%%%%%%%%%%%%%%%%%%%%%%%%%%%%%%%%%%%%%%%%%%%%%%%%%%%%%%%%%%%%%%%%%%%%%%%

The Sachdev-Ye-Kitaev (SYK) model
\cite{Sachdev, kitaev,Maldacena:2016hyu}
and its holographic dual Jackiw-Teitelboim (JT) gravity 
\cite{Jackiw:1984je,Teitelboim:1983ux,Almheiri:2014cka,
      Maldacena:2016upp,Jensen:2016pah,Engelsoy:2016xyb}
are useful testing ground to study various issues in quantum gravity
and holography. This duality is based on the fact that the 1d Schwarzian
theory, which arises from the Nambu-Goldstone mode of the spontaneously
broken time-reparametrization symmetry of the SYK model, also appears as
the boundary mode dynamics of JT gravity on asymptotic $AdS_2$.
This duality tells us that the random average of the
thermal partition function
$\bra Z(\bt)\ket=\bra\Tr e^{-\bt H_{\text{SYK}}}\ket$ of the SYK model 
reduces, at large number $N_{\text{SYK}}$ of fermions  and at low
energy, to the partition function of 
JT gravity on Euclidean $AdS_2$ which is topologically a disk
with renormalized boundary length $\bt$.

Recently, Saad, Shenker and Stanford \cite{Saad:2019lba} 
found that one can go beyond the strict large $N_{\text{SYK}}$
limit and actually compute the partition function of JT gravity
including the contribution of various topologies adding handles
(or Euclidean wormholes) to the disk.
They proposed that the partition function $Z_\JT(\bt)$ of JT gravity 
on asymptotically $AdS_2$ space is defined by a certain double-scaled
random matrix integral $\bra\Tr e^{-\bt H}\ket$,
where the Hamiltonian of the SYK model $H_{\text{SYK}}$
is replaced by a random hermitian matrix $H$.
Then the sum over topologies is reproduced from
the $1/N$ expansion of the matrix integral
with $N\sim e^{N_{\text{SYK}}}$.

Their proposal comes from the following facts \cite{Saad:2019lba}:
The path integral of JT gravity reduces to the contribution of the
Schwarzian mode describing the boundary wiggles, together with the 
Weil-Petersson volume $V_{g,1}(b)$ of the moduli space of Riemann
surfaces with $g$ handles and one geodesic boundary of length $b$.
The crucial point is that the recursion relation obeyed by the 
Weil-Petersson volume found by Mirzakhani \cite{Mirzakhani}
is equivalent to the topological recursion of
a double-scaled matrix model
with the spectral curve $y=\hf\sin(2z)$
\cite{Eynard:2007fi}.
Moreover, the genus-zero eigenvalue density $\rho_0(E)$
corresponding to this spectral curve is
exactly equal to the eigenvalue density
computed from the Schwarzian theory \cite{Stanford:2017thb}.
Since the topological recursion of Eynard and Orantin
\cite{Eynard:2007kz} is essentially determined by the data of spectral
curve (or $\rho_0(E)$) only, the above observations imply that the
boundary Schwarzian theory ``knows'' how to perform the sum over
topologies on the bulk JT gravity side.
It is further argued in \cite{Saad:2019lba} that this relation between
JT gravity and the matrix model
is generalized to arbitrary number of boundaries.

In this paper we will study the proposal in \cite{Saad:2019lba}
more closely, focusing on the single boundary case.
We find that the matrix model of JT gravity in \cite{Saad:2019lba}
is nothing but a special case of the old matrix model
of 2d gravity coupled to $c\leq1$ matter
\cite{Gross:1989vs,Gross:1989aw,Douglas:1989ve,Brezin:1990rb}
(see also \cite{Ginsparg:1993is} for a review). 
The important difference from the old story is that in the JT gravity
case infinitely many closed string couplings $t_n$
are turned on in a specific way:
$t_0=t_1=0,\ t_k=\tfrac{(-1)^k}{(k-1)!}\ (k\ge 2)$.\footnote{
It is advocated in \cite{Saad:2019lba} that
this background corresponds to
a $p\to\infty$ limit of the $(2,p)$ minimal string theory
\cite{Seiberg:2004at,Seiberg:2003nm}.}
We then introduce a natural two-parameter generalization
of $Z_\JT(\bt)$ by
releasing $t_0$ and $t_1$ from the above constraint.
Using this
we find that the partition function of JT gravity $Z_\JT(\bt)$
is written as the expectation value of the 
macroscopic loop operator $\Tr (e^{\bt Q}\Pi)$ \cite{Banks:1989df},
where $Q=\del_x^2+u(x)$ is the Lax operator and $\Pi$ 
is the projection to the states below the Fermi level. We will show that
this expression of $\Tr (e^{\bt Q}\Pi)$ naturally includes both of the
Schwarzian contribution and the Weil-Petersson volume.

This rewriting of the partition function using the Lax operator $Q$
is not just a formal expression, but is
very useful in practice for 
the actual computation of the genus expansion.
We will develop a systematic method of computing the higher genus
corrections to $Z_\JT$ using the Korteweg-de Vries (KdV)
equations obeyed by $u(x)$ and $\del_x Z_\JT$,
generalizing the approach of Zograf \cite{zograf1}.
Using this method we have computed the genus expansion of $Z_\JT$
up to $g=46$.
It turns out that this genus expansion of $Z_\JT$
is valid in the high temperature regime $(\bt\ll1)$.
In the low temperature regime, on the other hand,
we can compute 
$Z_\JT$ as a series expansion in $T=\bt^{-1}$
using the same KdV equations as above. 
We find that this low temperature expansion
can be rearranged 
by taking a scaling limit, which we will call the 't Hooft limit: 
\begin{equation}
\begin{aligned}
 \hbar\to0,~\bt\to\infty\quad \text{with}~\la=\hbar \bt~~\text{fixed},
\end{aligned} 
\label{eq:tHooft}
\end{equation}
where $\hbar=1/N$ is the genus counting parameter. 
It turns out that the free energy 
in the 't Hooft limit admits an open string like expansion
\begin{equation}
\begin{aligned}
 \cF=\log Z_\JT=\sum_{n=0}^\infty \hbar^{n-1}\cF_n(\la),
\end{aligned} 
\label{eq:thooft-free}
\end{equation}
and we find the first few terms of $\cF_n(\la)$
in a closed form.

Another interesting quantity to consider is the Baker-Akhiezer (BA)
function $\psi(E)$,
which is a solution of the Schr\"{o}dinger equation $-Q\psi(E)=E\psi(E)$
and interpreted as the wavefunction of FZZT brane 
\cite{Fateev:2000ik,Teschner:2000md}.
We find that the Laplace transform $\h{\psi}(\la)$ of the BA function
has a natural expansion in the 't Hooft limit.
We also study the behavior of the eigenvalue density $\rho(E)$ and
the BA function $\psi(E)$ by numerically evaluating the inverse Laplace
transform of $e^{\cF(\la)}$ and $\h{\psi}(\la)$.
We confirm the oscillating behavior of $\rho(E)$ and $\psi(E)$
in the classically allowed region $E>0$ discussed in \cite{Saad:2019lba}
which is non-perturbative in the coupling $\hbar$.

This paper is organized as follows.
In section \ref{sec:part}, we develop a technique of
the genus and the low temperature expansions
of $Z_\JT$ based on the KdV equations generalizing
the approach of \cite{zograf1}. Along the way,
we show that $Z_\JT$ is written as the expectation value 
of the macroscopic loop operator $\Tr(e^{\bt Q}\Pi)$.
In section \ref{sec:limit},
we consider the low energy expansion of $\rho(E)$ and $\psi(E)$,
as well as the corresponding low temperature expansion
of $Z_\JT$ and $\h{\psi}$ in the 't Hooft limit.
In section \ref{sec:num}, we study the behavior of $\rho(E)$ and
$\psi(E)$ numerically.
In section \ref{sec:SFF}, we comment on the connected correlator
$\bra Z(\bt_1)Z(\bt_2)\ket_{\text{conn}}$ and its analytic continuation
known as the spectral form factor.
Finally we conclude in section \ref{sec:discussion} with some
discussions for the interesting future directions.
In appendix \ref{app:airy} we summarize the known facts in the Airy case
described by the spectral curve $y=z$.
In appendix \ref{app:partial} we consider a partial resummation of the
genus expansion. In appendix \ref{app:string} we consider the so-called
string equation for the JT gravity case. In appendix \ref{app:BA}
we summarize useful properties of
the resolvent and the wave functions for the Schr\"odinger equation.

%%%%%%%%%%%%%%%%%%%%%%%%%%%%%%%%%%%%%%%%%%%%%%%%%%%%%%%%%%%%%%%%%%%%%%%%
\section{General properties of partition function\label{sec:part}}
%%%%%%%%%%%%%%%%%%%%%%%%%%%%%%%%%%%%%%%%%%%%%%%%%%%%%%%%%%%%%%%%%%%%%%%%

In this section we will show that JT gravity is
realized as the conventional 2d topological gravity
in the background where infinitely many couplings are turned on
in a specific way.
We will consider the partition function of JT gravity on
Riemann surfaces with one boundary and introduce
its two-parameter generalization.
The generalized partition function is closely related to
the tau-function for the KdV hierarchy.
Using this relation we will derive a simple differential equation
which uniquely determines the partition function
both in the genus and the low temperature expansions.

%%%
\subsection{JT gravity as 2d gravity in specific coupling background
\label{sec:Zbeta}}
%%%

Before discussing the partition function of JT gravity,
let us first recall some useful properties of
the partition function of the general 2d topological gravity
which we will use shortly.
(See e.g.~\cite{Dijkgraaf:2018vnm} for a recent review).
Let $\Sigma$ be a closed Riemann surface of genus $g$
with $n$ marked points $p_1,\ldots,p_n$
and let ${\cal M}_{g,n}$ be the moduli space of $\Sigma$.
We are interested in the intersection numbers
\begin{align}
\label{eq:intersec}
\langle\kappa^m\tau_{d_1}\cdots\tau_{d_n}\rangle
=\int_{\overline{\cal M}_{g,n}}
 \kappa^m\psi_1^{d_1}\cdots\psi_n^{d_n},\qquad
m,d_1,\ldots,d_n\in\bbZ_{\ge 0},
\end{align}
which are viewed as the correlation functions
of the 2d topological gravity.
Here, $\kappa$ (often denoted as $\kappa_1$ in the literature)
is the first Miller-Morita-Mumford
class and is proportional to the Weil-Petersson symplectic form
\begin{align}
\label{eq:omegakappa}
\omega=2\pi^2\kappa.
\end{align}
$\psi_i$ is the first Chern class of the complex line bundle
whose fiber is the cotangent space to $p_i$
and $\tau_{d_i}=\psi_i^{d_i}$.
$\overline{\cal M}_{g,n}$ is
the Deligne-Mumford compactification of
the moduli space ${\cal M}_{g,n}$.
Note that \eqref{eq:intersec} vanishes
unless $m+d_1+\cdots+d_n=3g-3+n$.

For the above correlation functions
one can introduce the formal generating function
\begin{align}
\label{eq:genG}
G(s,\{t_k\}):=\sum_{g=0}^\infty \gs^{2g}
 \left\langle e^{s\kappa+\sum_{d=0}^\infty t_d\tau_d}\right\rangle_g.
\end{align}
It is proved in \cite{mulase} (see also \cite{Dijkgraaf:2018vnm}) that
the intersection numbers involving both $\kappa$ and $\psi$'s
can be obtained from those involving $\psi$'s only.
More specifically, let $F$ be the formal generating function
that involves $\psi$'s only
\begin{align}
\label{eq:genF}
F(\{t_k\}):=\sum_{g=0}^\infty \gs^{2g}
 \left\langle e^{\sum_{d=0}^\infty t_d\tau_d}\right\rangle_g.
\end{align}
$G$ is then given by
\begin{align}
G(s,\{t_k\})=F(\{t_k+\gamma_k s^{k-1}\})
\end{align}
with
\begin{align}
\label{eq:gammavalue1}
\gamma_0=\gamma_1=0,\quad
\gamma_k=\frac{(-1)^k}{(k-1)!}
\quad (k\ge 2).
\end{align}
By using this property
we will see that JT gravity is
nothing but the special case of
the topological gravity with $t_k=\gamma_k$. 

Let us now consider
the partition function of JT gravity
on two-dimensional surfaces of arbitrary genus
with one boundary.
In \cite{Saad:2019lba} this partition function is evaluated as
the one-point correlation function
\begin{align}
\langle Z(\beta)\rangle
 =\langle Z(\beta)\rangle_{g=0}
 +\langle Z(\beta)\rangle_{g\ge 1},
\end{align}
where $Z(\beta)=\Tr e^{-\bt H}$ is the thermal partition function of
a certain Hermitian matrix model.
The genus-zero part is to be evaluated separately.
Let us first begin with the $g\ge 1$ part.
In \cite{Saad:2019lba} it is evaluated as\footnote{
Throughout this paper we fix the normalization of the
Weil-Petersson form by setting $\alpha=1$ (see \cite{Saad:2019lba}).}
\begin{align}
\label{eq:Zg1a}
\begin{aligned}
\langle Z(\beta)\rangle_{g\ge 1}
&=\sum_{g=1}^\infty e^{(1-2g)S_0}
 \int_0^\infty bdb Z_\Sch^\trumpet(\beta,b) V_{g,1}(b)\\
&=\sum_{g=1}^\infty e^{(1-2g)S_0}
  \int_0^\infty bdb 
  \frac{\gamma^{1/2}e^{-\frac{\gamma b^2}{2\beta}}}
       {(2\pi\beta)^{1/2}} V_{g,1}(b).
\end{aligned}
\end{align}
Here, $V_{g,1}(b)$ is the Weil-Petersson volume of the moduli space
of a genus $g$ surface with one geodesic boundary of length $b$
and $Z_\Sch^\trumpet(\beta,b)$ comes from the path integral of the
Schwarzian mode on the ``trumpet'' geometry.
$V_{g,1}(b)$ is given by
\begin{align}
V_{g,1}(b)
=\int_{\overline{\cal M}_{g,1}}
  \exp\left(2\pi^2\kappa+\frac{b^2}{2}\psi_1\right)
\equiv\left\langle
\exp\left(2\pi^2\kappa+\frac{b^2}{2}\psi_1\right)\right\rangle_{g,1}.
\end{align}
As mentioned below \eqref{eq:omegakappa},
the correlation function
$\langle \kappa^k\psi_1^l \rangle_{g,1}$ $(k,l\in\bbZ_{\ge 0})$
has the following property
\begin{align}
\label{eq:vcond}
\langle \kappa^k\psi_1^l \rangle_{g,1}=0
\quad\mbox{unless}\quad k+l=3g-2.
\end{align}
One can thus expand $V_{g,1}$ as
\begin{align}
V_{g,1}(b)
=\sum_{d=0}^{3g-2}
  \frac{(2\pi^2)^{3g-2-d}(b^2/2)^d}{(3g-2-d)!d!}
  \langle\kappa^{3g-2-d}\psi_1^d\rangle_{g,1}.
\label{eq:Vg1}
\end{align}
By plugging this expression into \eqref{eq:Zg1a}
and evaluating the integral, one obtains
\begin{align}
\label{eq:Zg1b}
\begin{aligned}
\langle Z(\beta)\rangle_{g\ge 1}
&=\frac{1}{\sqrt{2\pi}}\sum_{g=1}^\infty \gs^{2g-1}
  \sum_{d=0}^{3g-2}
  \frac{1}{(3g-2-d)!}
  \left(\frac{\beta}{2\pi^2\gamma}\right)^{d+1/2}
  \langle\kappa^{3g-2-d}\psi_1^d\rangle_{g,1},
\end{aligned}
\end{align}
where we have identified the genus counting parameter as
\begin{align}
\gs=(2\pi^2)^{3/2}e^{-S_0}.
\end{align}
On the other hand,
the genus-zero part comes from
the path integral of the Schwarzian mode on the disk,
which 
is expressed as \cite{Saad:2019lba}
\begin{align}
\label{eq:Zg0}
\langle Z(\beta)\rangle_{g=0}
=e^{S_0}Z_\Sch^\disk
=e^{S_0}
 \frac{\gamma^{3/2}e^{\frac{2\pi^2\gamma}{\beta}}}
      {(2\pi)^{1/2}\beta^{3/2}}
=\frac{1}{\sqrt{2\pi}}
 \gs^{-1}\left(\frac{2\pi^2\gamma}{\beta}\right)^{3/2}
 e^{\frac{2\pi^2\gamma}{\beta}}.
\end{align}
We see from \eqref{eq:Zg1b} and \eqref{eq:Zg0} that
it is convenient to absorb $\gamma$ into the normalization of $\beta$,
\begin{align}
\frac{\beta}{2\pi^2\gamma}\to \beta,
\end{align}
or equivalently, one can simply set $\gamma=1/2\pi^2$.
Doing this, we find
\begin{align}
\begin{aligned}
\langle Z(\beta)\rangle_{g=0}
&=\frac{1}{\sqrt{2\pi}}
 \gs^{-1}\beta^{-3/2}
 e^{\beta^{-1}},\\
\langle Z(\beta)\rangle_{g\ge 1}
&=\frac{1}{\sqrt{2\pi}}\sum_{g=1}^\infty\sum_{d=0}^\infty
  \gs^{2g-1}
  \beta^{d+1/2}
  \langle e^\kappa\psi_1^d\rangle_{g,1}.
\end{aligned}
\label{eq:ZJT-g}
\end{align}
The last expression is obtained from \eqref{eq:Zg1b}
with the help of the property \eqref{eq:vcond}.
Putting these expressions together we obtain
\begin{align}
\label{eq:ZJTresult1}
\langle Z(\beta)\rangle
 =
\frac{1}{\sqrt{2\pi}\gs\beta^{3/2}}\biggl(
e^{\beta^{-1}}
 +\sum_{g=1}^\infty\sum_{d=0}^\infty
  \gs^{2g}
  \beta^{d+2}
  \langle e^\kappa\psi_1^d\rangle_{g,1}\biggr).
\end{align}
From \eqref{eq:genG}--\eqref{eq:gammavalue1}
we see that
\begin{align}
\begin{aligned}
\sum_{g=1}^\infty \gs^{2g}\langle e^\kappa\psi_1^d\rangle_{g,1}
&=\partial_d G(s=1,\{t_k=0\})\\
&=\partial_d F\left(\{t_k=\gamma_k\}\right).
\end{aligned}
\end{align}
Plugging this into \eqref{eq:ZJTresult1} we finally obtain
\begin{align}
\label{eq:ZJTresult2}
\langle Z(\beta)\rangle
 =
\frac{1}{\sqrt{2\pi}\gs\beta^{3/2}}\biggl(
e^{\beta^{-1}}+\sum_{d=0}^\infty\beta^{d+2}
  \partial_d F\left(\{t_k=\gamma_k\}\right)\biggr).
\end{align}
We have thus shown that
the partition function of JT gravity
on surfaces with one boundary
is expressed entirely in terms of the general topological gravity
in a specific background $t_k=\gamma_k$.

%%%
\subsection{Generalized partition function and KdV constraints}
%%%

The relation \eqref{eq:ZJTresult2} of
JT gravity with the general topological gravity
provides us with a better understanding of
$\langle Z(\beta)\rangle$
as well as an efficient algorithm of computing it.
It is well known that the partition function of
the topological gravity obeys the KdV constraints
\cite{Witten,Kontsevich,Itzykson:1992ya}.
In fact, Zograf proposed an efficient algorithm of computing
the Weil-Petersson volume by making use of the KdV equation
\cite{zograf1}.
In what follows we will generalize his idea and
present a more direct application of the KdV constraints
to JT gravity.

Let us first recall how the KdV constraints occur
in the general topological gravity.
It was conjectured by Witten \cite{Witten}
and proved by Kontsevich \cite{Kontsevich}
(see also \cite{Itzykson:1992ya})
that $e^F$ with $F$ defined in \eqref{eq:genF}
is a tau function for the KdV hierarchy.
This means that
\begin{align}
u:=\partial_0^2 F
\end{align}
satisfies the (generalized) KdV equations
\begin{align}
\label{eq:KdVeqs}
\partial_k u = \partial_0\cR_{k+1},
\end{align}
where $\cR_k$ are the Gelfand-Dikii differential polynomials of $u$
\begin{align}
\label{eq:cRforms}
\cR_0=1,\quad
\cR_1=u,\quad
\cR_2=\frac{u^2}{2}+\frac{\Dpartial_0^2 u}{12},\quad
\cR_3=\frac{u^3}{6}+\frac{u\Dpartial_0^2 u}{12}
 +\frac{(\Dpartial_0 u)^2}{24}+\frac{\Dpartial_0^4 u}{240},\quad
\cdots.
\end{align}
Here we have introduced the notations
\begin{align}
\label{eq:xdef}
\partial_k :=\frac{\partial}{\partial t_k},\quad
\Dpartial_k :=\gs\partial_k.
\end{align}
For $k=1$, \eqref{eq:KdVeqs} gives the traditional KdV equation
\begin{align}
\label{eq:dlKdV}
\Dpartial_1 u
 =\Dpartial_0\left(\frac{u^2}{2}+\frac{\Dpartial_0^2 u}{12}\right).
\end{align}
$\cR_k$ are determined by the recursion relation
\begin{align}
\label{eq:rec_R}
(2k+1)\Dpartial_0 \cR_{k+1}
 =\frac{1}{4}\Dpartial_0^3\cR_k+2u\Dpartial_0\cR_k+(\Dpartial_0 u)\cR_k
\end{align}
with the initial condition $\cR_0=1$.
Integrating \eqref{eq:KdVeqs} once in $t_0$ we have
\begin{align}
\label{eq:FRrel}
\partial_k\partial_0 F = \cR_{k+1}.
\end{align}
In this paper we call the above relations obeyed by $F$
the KdV constraints.

We would like to make use of the KdV constraints
to study the JT gravity partition function \eqref{eq:ZJTresult2}.
To do this, it is better not to fix the value of $t_i$ completely 
as in \eqref{eq:ZJTresult2} but rather leave
$t_0$ and $t_1$ as parameters. In what follows we regard
$F$ as a function in $t_0,t_1$ (and also in $\gs$)
\begin{align}
F(t_0,t_1)=F\left(t_0,t_1,\{t_k=\gamma_k\}_{k\ge 2}\right).
\end{align}
As we will see, at least locally around $(t_0,t_1)=(0,0)$
one can introduce such a two-parameter deformation.
One should also keep in mind that there is no guarantee that
$F(t_0,t_1)$ is well-defined for arbitrary values of $(t_0,t_1)$.
For our purposes it is convenient to introduce the rescaled parameters
\begin{align}
\label{eq:rescaledvar}
\hbar:=\frac{1}{\sqrt{2}}\gs,\quad
x :=\hbar^{-1}t_0,\quad
\tau :=\hbar^{-1}t_1
\end{align}
and the notation
\begin{align}
\label{eq:rescaleddiff}
{}':=\partial_x=\hbar\partial_0,\quad
\dot{~}:=\partial_\tau=\hbar\partial_1.
\end{align}
We then introduce a two-parameter deformation
of the partition function \eqref{eq:ZJTresult2} as
\begin{align}
\label{eq:cZJTdef}
Z_\JT(t_0,t_1)
 :=
\frac{1}{\sqrt{2\pi}\gs\beta^{3/2}}\biggl(
   e^{\beta^{-1}}+\beta t_0
  +\sum_{k=0}^\infty\beta^{k+2}\partial_k F(t_0,t_1)\biggr).
\end{align}
$Z_\JT$ reproduces $\langle Z(\beta)\rangle$ as
\begin{align}
Z_\JT(0,0)=\langle Z(\beta)\rangle.
\end{align}
We have added the term $\beta t_0$
in the definition of $Z_\JT(t_0,t_1)$ in \eqref{eq:cZJTdef}
so that we obtain a simple relation
\begin{align}
\label{eq:Wdef}
\begin{aligned}
\partial_x Z_\JT
 &= \frac{1}{2\sqrt{\pi\beta}}\left(
1+\sum_{k=0}^\infty \beta^{k+1}\partial_k\partial_0 F\right)\\
 &=\frac{1}{2\sqrt{\pi\beta}}\sum_{k=0}^\infty \beta^k \cR_k\\
 &=: W,
\end{aligned}
\end{align}
where we have used $\cR_0=1$ and \eqref{eq:FRrel}.
$Z_\JT$ is thus computed from
the generating function $W$ for
the Gelfand-Dikii polynomials $\cR_k$.

The Laplace transform of $W$
\begin{align}
R(\xi)
 =\int_0^\infty d\beta e^{-\beta\xi}W(\beta)
\end{align}
has a beautiful interpretation. It is expanded as
\begin{align}
R = \sum_{k=0}^\infty \xi^{-k-1/2} R_k
\end{align}
with coefficients being again the Gelfand-Dikii polynomials
\begin{align}
R_k = \frac{(2k-1)!!}{2^{k+1}}\cR_k.
\label{eq:R-to-cR}
\end{align}
In this notation $R_k$ are written as
\begin{align}
R_0=\frac{1}{2},\quad
R_1=\frac{u}{4},\quad
R_2=\frac{1}{16}(3u^2+u''),\quad
R_3=\frac{1}{64}(10u^3+10uu''+5{u'}^2+u''''),\quad
\cdots.
\end{align}
With change of notation $u\to -u$ these $R_k$ are
identified precisely with the original polynomials appeared
in the paper of Gelfand and Dikii \cite{Dikii}.
This means that their generating function $R$
is the resolvent \cite{Dikii}
\begin{align}
\label{eq:resolvent}
R(\xi)=\Bigl\langle x\,\Big|\,\frac{1}{\xi-Q}\,\Big|\,x\Bigr\rangle
\end{align}
for the Schr\"odinger equation
\begin{align}
Q\psi=\xi\psi
\end{align}
with
\begin{align}
Q:=\partial_x^2 +u.
\label{eq:Q}
\end{align}
Here, $|x\rangle$ is the coordinate eigenstate.
Note that $Q$ is nothing but the Lax operator $L$
for the KdV equation, which we will discuss later.

By taking the inverse Laplace transform of \eqref{eq:resolvent} 
we obtain the formal expression
\begin{align}
\label{eq:Wformula}
W = \langle x|e^{\beta Q}|x\rangle.
\end{align}
From the relation $\del_x Z_\JT=W$ in \eqref{eq:Wdef}, 
we find 
\begin{align}
Z_\JT = \int_{-\infty}^x dx' \langle x'|e^{\beta Q}|x'\rangle.
\label{eq:macro-loop}
\end{align}
Introducing the projector $\Pi$ by
\begin{equation}
\begin{aligned}
 \Pi=\int_{-\infty}^x dx'|x'\ket\bra x'|,
\end{aligned} 
\label{eq:Pi}
\end{equation}
we arrive at a very simple expression of $Z_\JT$
\begin{equation}
\begin{aligned}
 Z_\JT=\Tr(e^{\beta Q}\Pi).
\end{aligned} 
\label{eq:macro-Pi}
\end{equation}
Topological gravity and other models of 2d gravity coupled to matter
are described by a double-scaling limit of the general matrix model,
in which $\Tr (e^{\beta Q}\Pi)$ is known as (the expectation value of)
the macroscopic loop operator \cite{Banks:1989df}.\footnote{
Our definition of the sign of $x$ is opposite from that in
\cite{Banks:1989df}; in \cite{Banks:1989df} the projector is given by
$\Pi=\int_{x}^\infty dx'|x'\ket\bra x'|$ while in our definition $\Pi$
is given by \eqref{eq:Pi}.}
We have thus shown that the partition function of JT gravity
on surfaces with one boundary is identified with
a single macroscopic loop operator of the matrix model.
In this sense JT gravity is merely an example of the old 2d gravity
(see \cite{Ginsparg:1993is} for a review).
What is special about JT gravity, when compared with the previously
known examples, is that infinitely many couplings $t_n$ are turned
on with a specific value $t_n=\ga_n$ in \eqref{eq:gammavalue1}.

%%%
\subsection{Lax formalism and master differential equation }
%%%

As is well known, the KdV equation admits the Lax formalism.
This enables us to derive a simple differential equation
for $W$, which can be used to compute $Z_\JT$.

A crucial fact about the resolvent $R$ is that
it is written as \cite{DiFrancesco:1993cyw}
(see also appendix \ref{app:BA})
\begin{align}
\label{eq:Rinpsi}
R=\psi_+\psi_-,
\end{align}
where $\psi_\pm$
are certain two independent solutions to the auxiliary linear problem
\begin{align}
\label{eq:linprob}
L\psi_a=\xi\psi_a,\quad
\dot{\psi}_a=M\psi_a.
\end{align}
Here
\begin{align}
\begin{aligned}
L=Q&=\partial_x^2+u,\qquad
M=\frac{2}{3}\partial_x^3+u\partial_x+\frac{1}{2}u'
\end{aligned}
\end{align}
are the Lax pair for the KdV equation.
In fact, in the rescaled notation
\eqref{eq:rescaledvar}--\eqref{eq:rescaleddiff}
the KdV equation \eqref{eq:dlKdV} is written as
\begin{align}
\label{eq:dlKdV2}
\dot{u}=\frac{1}{6}u'''+uu'
\end{align}
and it is obtained as the compatibility condition
\begin{align}
\dot{L} = [M,L]
\end{align}
for the linear problem \eqref{eq:linprob}.

From \eqref{eq:Rinpsi}--\eqref{eq:linprob} one can show that
\begin{align}
\begin{aligned}
\xi R'
 &=\frac{1}{4}R'''+uR'+\frac{1}{2}u'R,\\
\dot{R}
 &=\frac{1}{6}R'''+uR'.
\end{aligned}
\label{eq:R-eq}
\end{align}
The first equation is equivalent to
the recursion relation \eqref{eq:rec_R},
which is written for $R_k$ as
\begin{align}
R_{k+1}'=\frac{1}{4}R_k'''+uR_k'+\frac{1}{2}u'R_k.
\end{align}
From the second equation it immediately follows that
\begin{align}
\label{eq:Wdiffeq}
\dot{W} = \frac{1}{6}W'''+u W'.
\end{align}
We have thus derived a simple, linear differential equation
for $W=\partial_x Z_\JT$.
Explicitly in terms of $Z_\JT$ it is expressed as
\begin{align}
\partial_\tau\partial_x Z_\JT
=\frac{1}{6}\partial_x^4 Z_\JT + u\partial_x^2 Z_\JT.
\end{align}
%

%%%
\subsection{Genus expansion}\label{sec:genus}
%%%

We can use the differential equation \eqref{eq:Wdiffeq}
together with the KdV equation \eqref{eq:dlKdV2} to compute
$Z_\JT$ as a power series expansion in $\gs$.
For this purpose, it is convenient to rewrite 
these equations
in such a way that the $\gs$-dependence is manifest
\begin{align}
\label{eq:gsKdV}
\partial_1 u&=u\partial_0 u +\frac{\gs^2}{12}\partial_0^3u,\\
\label{eq:W-eq}
\partial_1 W&=u\partial_0 W +\frac{\gs^2}{12}\partial_0^3W.
\end{align}
As we will see below, the genus expansion of $u$ and $W$
are completely determined by these equations.
Prior to the practical computation
it is useful to recall the following fact \cite{Itzykson:1992ya}:
$F_g=\langle e^{\sum_{d=0}^\infty t_d\tau_d}\rangle_g\ (g\ge 2)$
is a polynomial in $I_k\ (k\ge 2)$ and $(1-I_1)^{-1}$,
where\footnote{The coupling $t_k$ and the parameter
$\La=\diag(\la_1,\cdots,\la_{M})$
in the Kontsevich model \cite{Kontsevich}
are related by the so-called Miwa transformation
\begin{equation}
\begin{aligned}
 t_k=-(2k-1)!!\Tr\La^{-2k-1}.
\end{aligned} 
\label{eq:Miwa}
\end{equation}
In terms of $\La$, $I_k$ is written as
\cite{Makeenko:1991ec,Ambjorn:1993sj}
\begin{equation}
\begin{aligned}
 I_k=
-(2k-1)!!\Tr (\La^2-2u_0)^{-k-\hf}.
\end{aligned} 
\end{equation}
It is interesting to observe that going from $t_k$ to $I_k$
amounts to shifting $\La^2\to\La^2-2u_0$.
Note that $\ga_k$ in \eqref{eq:gammavalue1} is written as
a contour integral on the spectral curve
$y=\hf \sin(2\rt{\xi})$ \eqref{eq:curve}
\begin{equation}
\begin{aligned}
 \ga_k=-(2k-1)!!2^{1-k}\oint\frac{d\xi}{2\pi\ri}y(\xi)\xi^{-k-\hf}.
\end{aligned} 
\end{equation}
\label{footnote}
}
\begin{align}
I_k(u_0,\{t_k\}):=\sum_{n=0}^\infty t_{n+k}\frac{u_0^n}{n!}
\label{eq:Ik-def}
\end{align}
with $u_0:=\partial_0^2 F_0$.
In our present case with $t_k=\gamma_k\ (k\ge 2)$,
where $\gamma_k$ is given in \eqref{eq:gammavalue1},
it is convenient to introduce the new variables
\begin{align}
y:=u_0,\quad
t:=1-I_1
\label{eq:yt-def}
\end{align}
and the functions
\begin{align}
\label{eq:Bndef}
B_n(y)
 :=\frac{J_n(2\sqrt{y})}{y^{n/2}}
  =\sum_{k=\max(0,-n)}^\infty\frac{(-1)^ky^k}{k!(k+n)!}\qquad(n\in\bbZ).
\end{align}
Here, $J_n(z)$ is the Bessel function of the first kind.
One then finds that
$I_n\ (n\ge 2)$ are identified as
\begin{align}
I_n(y,\{t_0,t_1,t_k=\gamma_k\ (k\ge 2)\})
=(-1)^nB_{n-1}.
\end{align}
Therefore, in our case
$F_g\ (g\ge 2)$ is a polynomial in $B_n(y)\ (n\ge 1)$ and $t^{-1}$.
Note that $B_n$ satisfies
\begin{align}
\partial_yB_n=-B_{n+1},\qquad
yB_{n+1}=nB_n-B_{n-1}
\end{align}
and also
\begin{align}
B_n(0)=\frac{1}{n!},\quad n\ge 0.
\end{align}
The old variables $(t_0,t_1)$ and the new ones $(y,t)$
are related as
\begin{align}
t_1=B_0-t,\quad
t_0=y(B_1-t_1).
\label{eq:t1t0}
\end{align}
The first equation of \eqref{eq:t1t0} simply follows from
the definition of $I_1$ in \eqref{eq:Ik-def},
while the second equation of \eqref{eq:t1t0} comes from
the classical, $\hbar\to0$ limit of the string equation 
\eqref{eq:qu-string} \cite{Itzykson:1992ya}
\begin{equation}
\begin{aligned}
 u_0-I_0(u_0,\{t_k\})=0.
\end{aligned} 
\label{eq:classical-string}
\end{equation}
This relation \eqref{eq:classical-string}
can be interpreted as the stationarity condition
$\frac{\del F_0}{\del u_0}=0$
of the genus-zero free energy \cite{Itzykson:1992ya}
\begin{equation}
\begin{aligned}
 F_0=\hf\int_0^{u_0}du \bigl[u-I_0(u,\{t_k\})\bigr]^2.
\end{aligned} 
\end{equation}
In terms of $(y,t)$ the differentials $\partial_{0,1}$ are written as
\begin{align}
\partial_0=\frac{1}{t}(\partial_y-B_1\partial_t),\quad
\partial_1=y\partial_0-\partial_t.
\end{align}
This is identical to the change of variables introduced by Zograf
(see \cite{zograf1}).
One can show that the ``on-shell'' value 
$(t_0,t_1)=(0,0)$ corresponds to $(y,t)=(0,1)$.
It is important to note that the map \eqref{eq:t1t0}
from $(y,t)$ to $(t_0,t_1)$ is
neither one-to-one nor onto.
This gives rise to a somewhat delicate issue:
$Z_\JT$ and related quantities we are going to solve
become multivalued and not globally well-defined
(at least as real functions)
when viewed as functions in $(t_0,t_1)$.
But as far as local expansions around the ``on-shell'' value
are concerned, one can ignore such intricacies.

With these preparations let us first consider the genus expansion
of $u$. By construction $u$ has the expansion of the form
\begin{align}
\label{eq:ugsexp}
u=\sum_{g=0}^\infty \gs^{2g}u_g
\end{align}
with $u_g=\partial_0^2F_g$.
One can easily show that
the differential equation \eqref{eq:gsKdV}
is written as the recursion relation
\begin{align}
\label{eq:urecrel}
-\frac{1}{t}\partial_t(tu_g)
=\sum_{h=1}^{g-1}u_{g-h}\partial_0 u_h
 +\frac{1}{12}\partial_0^3u_{g-1}
\quad (g\ge 1).
\end{align}
Then $u_g$ are obtained by recursively solving this equation
with the initial condition $u_0=y$.
First two of them are
\begin{align}
\begin{aligned}
u_1&=\frac{B_1^2}{12t^4}-\frac{B_2}{24t^3},\\
u_2&=\frac{49B_1^5}{288t^9}-\frac{11B_1^3B_2}{36t^8}
 +\frac{84B_1^2B_3+109B_1B_2^2}{1152t^7}
 -\frac{32B_1B_4+51B_2B_3}{2880t^6}
 +\frac{B_5}{1152t^5}.
\end{aligned}
\end{align}
Precisely speaking, the equation \eqref{eq:urecrel} by itself
does not determine the ``integration constant,''
i.e.~there is freedom to add a term linear in $t^{-1}$
at each step of the recursion.
Such a term is, however, forbidden
by the relation $u_g=\partial_0^2 F_g$.
For instance, the above results of $u_g$
can also be obtained from
the results of $F_g\ (g=1,2)$ in \cite{Itzykson:1992ya}
\begin{align}
\begin{aligned}
F_1&=-\frac{1}{24}\ln t,\\
F_2&=\frac{7B_1^3}{1440t^5}-\frac{29B_1B_2}{5760t^4}
  +\frac{B_3}{1152t^3}
\end{aligned}
\end{align}
and one can verify that
$u_g\ (g=1,2)$ do not contain any terms linear in $t^{-1}$.
More generally, since $F_g\ (g\ge 2)$ are polynomials in $t^{-1}$
and the action of $\partial_0$ increases the degree of $t^{-1}$
at least by one,
$u_g\ (g\ge 2)$ cannot have any terms linear in $t^{-1}$.
Therefore, one can in fact unambiguously determine $u_g$
by recursively solving \eqref{eq:urecrel}.
Note that the polynomial structure of $F_g$ also allows us
to compute it unambiguously from $u_g$.

Let us next consider the genus expansion of $W$.
In contrast to $u$,
$W$ depends not only on $y,t,\gs$ but also on $\beta$,
though $\beta$ does not appear explicitly
in the differential equation \eqref{eq:Wdiffeq}.
The genus-zero part of $W$ is obtained
from \eqref{eq:Wformula} by ignoring the commutator
of $\partial_x$ and $u(x)$ and by performing the integral
with respect to the momentum $p=\ri^{-1}\partial_x$
\begin{align}
W_{g=0}
 =\int_{-\infty}^\infty\frac{dp}{2\pi}e^{\beta (-p^2+u_0)}
 =\frac{e^{\beta u_0}}{2\sqrt{\pi\beta}}.
\label{eq:Wg=0}
\end{align}
Alternatively, this form can also be obtained from \eqref{eq:Wdef}
by noticing the following property of the Gelfand-Dikii polynomials
\begin{align}
\label{eq:cRprop}
\cR_k
 =\frac{u^k}{k!}+{\cal O}(D_0^2)
 =\frac{u_0^k}{k!} +{\cal O}(\gs^2).
\end{align}
Let us then expand $W$ as
\begin{align}
\label{eq:Wgsexp}
W=\frac{e^{\beta y}}{2\sqrt{\pi\beta}}\sum_{g=0}^\infty\gs^{2g}W_g.
\end{align}
We have chosen the overall factor so that $W_0=1$.
By plugging the genus expansions \eqref{eq:Wgsexp} and \eqref{eq:ugsexp}
into the differential equation \eqref{eq:Wdiffeq}
we obtain the recursion relation
\begin{align}
\label{eq:wrecrel}
-\partial_t W_g
 =\sum_{h=0}^{g-1} u_{g-h}\tpartial_0 W_h
 +\frac{1}{12}\tpartial_0^3W_{g-1}\quad(g\ge 1),
\end{align}
where $\tpartial_0$ is given by
\begin{align}
\tpartial_0=e^{-\beta y}\partial_0 e^{\beta y}=\partial_0 +\beta t^{-1}.
\label{eq:tilde-del0}
\end{align}
The equation \eqref{eq:wrecrel} by itself again does not determine
the $t$-independent part of $W_g$.
However, if we express $W$ in terms of $u$ using
\eqref{eq:Wdef} and \eqref{eq:cRforms}
and consider the genus expansion,
we see that the only possible source of $t$-independent term is
$u_0$ with no derivatives acting on it.
As we saw above, the contribution of such $u_0$
is entirely captured by the overall factor in \eqref{eq:Wgsexp}
and consequently $W_g\ (g\ge 1)$ does not contain
any $t$-independent term.
We can therefore unambiguously compute $W_g$ by recursively solving
\eqref{eq:wrecrel}, starting from the initial condition $W_0=1$.
For instance, we find
\begin{align}
\begin{aligned}
W_1&=\frac{\beta^3}{24t^2}+\frac{2\beta^2B_1-\beta B_2}{24t^3}
  +\frac{\beta B_1^2}{12t^4}.
\end{aligned}
\label{eq:W1}
\end{align}
It is also easy to prove that
$W_g$ is a polynomial of weight $(3g,-2g)$
in the generators $B_n\ (n\ge 1)$, $\bt$ and $t^{-1}$,
to which weights $(n,1)$, $(1,0)$ and $(0,-1)$
are assigned respectively.

Finally let us consider the genus expansion of $Z_\JT$
\begin{align}
Z_\JT=\frac{e^{\beta y}}{\sqrt{2\pi\beta^3}\gs}
 \sum_{g=0}^\infty \gs^{2g}Z_g.
\label{eq:ZJT-expand}
\end{align}
From the relation $\del_xZ_\JT=W$, one can show that 
$Z_g$ and $W_g$ are related by
\begin{align}
\beta^{-1}\tilde{\del}_0Z_g=W_g,
\label{eq:Wg-Zg}
\end{align}
where $\tpartial_0$ is defined in \eqref{eq:tilde-del0}.
Note that the extra factor $\bt^{-1}$ comes from the
difference of the powers of $\bt$ in the prefactor of $W$
\eqref{eq:Wgsexp} and $Z_\JT$ \eqref{eq:ZJT-expand}.
One can prove that $W_g\ (g\ge 1)$ has the structure
$W_g=\sum_{k=2g}^{5g-1}(t^{-1})^k W_g^{(k)}$
when written as a polynomial in $t^{-1}$.
It then follows from \eqref{eq:Wg-Zg}
that $Z_g\ (g\ge 1)$ has the structure
$Z_g=\sum_{k=2g-1}^{5g-3}(t^{-1})^k Z_g^{(k)}$.
Then using \eqref{eq:Wg-Zg} we can easily compute
$Z_g\ (g\ge 1)$ from the result of $W_g$
by iteratively determining the coefficient $Z_g^{(k)}$
in descending order with respect to $k$.
For example, from the result of $W_1$ in \eqref{eq:W1}
we find
\begin{align}
\begin{aligned}
Z_1&=\frac{\beta^3}{24t}+\frac{\beta^2 B_1}{24t^2}.
\end{aligned}
\end{align}
We have checked that the on-shell value of $Z_g$ reproduces the known
intersection numbers 
\begin{align}
Z_g\Big|_{t=1,y=0}
 =\sum_{d=0}^{3g-2}\beta^{d+2}\langle e^\kappa\psi_1^d\rangle_{g,1}.
\end{align}

One can see that $Z_g$ becomes small when $\bt\ll1$,
i.e.~at high temperature $T=\bt^{-1}\gg1$.
In this sense the genus expansion of $Z_\JT$ in \eqref{eq:ZJT-expand}
can be thought of as a high temperature expansion.
As we will see in section \ref{sec:low}
we can consider the opposite low temperature limit $T\ll1$.
To study the low temperature regime it is useful to define
\begin{equation}
\begin{aligned}
 \til{Z}_g=\bt^{-3g}Z_g\Big|_{y=0,t=1}
=\sum_{\ell=0}^{3g-2-\ell}\frac{T^\ell}{\ell!}
 \langle \kappa^\ell\psi_1^{3g-2-\ell}\rangle_{g,1}
=\Biggl\bra \frac{e^{T\ka}}{1-\psi_1}\Biggr\ket_{g,1}.
\end{aligned} 
\end{equation}
In the last equality we have used the selection rule \eqref{eq:vcond}.
The first three terms are 
\begin{equation}
\begin{aligned}
\til{Z}_1&=\frac{1}{24}+\frac{T}{24},\\
\til{Z}_2&=\frac{1}{1152}+\frac{29T}{5760}+\frac{139T^2}{11520}
 +\frac{169T^3}{11520}+\frac{29T^4}{3072},\\
\til{Z}_3&=\frac{1}{82944}+\frac{77T}{414720}+\frac{3781T^2}{2903040}
 +\frac{47209T^3}{8709120}+\frac{127189T^4}{8709120}\\
&\quad+\frac{8983379T^5}{348364800}+\frac{8497697T^6}{298598400}
+\frac{9292841 T^7}{522547200}.
\end{aligned} 
\end{equation}
Using the above algorithm we have computed $\til{Z}_g$
up to $g=46$.\footnote{
The data of $\til{Z}_g (g=1,\cdots,46)$ are attached to
the arXiv submission in the file \texttt{zdata.txt}.
The reader can import this file to \texttt{Mathematica} by the command
 \texttt{data=Get["./zdata.txt"];}.
Then $\texttt{data[[g]]}$ returns  $\til{Z}_g$.
We have checked that our data agree with the result of Zograf
up to $g=20$ \cite{zograf-data}.
}
These data provide us with valuable information of 
the large genus behavior of the genus expansion.
In particular, we find the all-genus result of the intersection number
$\langle \kappa^\ell\psi_1^{3g-2-\ell}\rangle_{g,1}$
with fixed $\ell$ 
\begin{equation}
\begin{aligned}
 \langle \kappa^\ell\psi_1^{3g-2-\ell}\rangle_{g,1}
 =\frac{P_\ell(g)}{(24)^gg!},
\end{aligned} 
\label{eq:Pell-def}
\end{equation}
where $P_\ell(g)$ is a degree-$2\ell$ polynomial of $g$.
The $\ell=0$ case is computed in \cite{Itzykson:1992ya} with the famous
result $P_0(g)=1$.
The $\ell=1$ case has appeared in \cite{BDY} with the result
\begin{equation}
\begin{aligned}
 P_1(g)=1+\frac{12}{5}g(g-1).
\end{aligned} 
\end{equation}
From our data of $\til{Z}_g$ we find $P_\ell(g)$ for $\ell\geq2$,
which are not known in the literature:  
\begin{equation}
\begin{aligned}
P_2(g)&=\frac{1}{175} (g-1) \left(1008
   g^3-1200 g^2+888
   g-175\right),\\
P_3(g)&=\frac{1}{875} (g-1)
   \left(12096 g^5-31104
   g^4+35856 g^3-25644
   g^2+7960 g-875\right),\\
P_4(g)&=\frac{1}{336875}(g-1)(11176704 g^7-46303488
   g^6+82114560 g^5-89621280
   g^4\\
&\qquad \qquad +62820096 g^3-22974252
   g^2+8338585
   g-336875).
\end{aligned} 
\end{equation}
One can see from \eqref{eq:Pell-def} that
$\langle \kappa^\ell\psi_1^{3g-2-\ell}\rangle_{g,1}$
does not exhibit the usual $(2g)!$ growth with fixed $\ell$.
The $(2g)!$ growth comes from the opposite end
$\langle \kappa^{3g-2-d}\psi_1^{d}\rangle_{g,1}$
with fixed small $d$ \cite{zograf1,zograf2}.
One can also see that the sum over genus of \eqref{eq:Pell-def}
is convergent, which we will study in detail in section \ref{sec:low}.

From our data of $\til{Z}_g$  up to $g=46$,
we have extracted numerically the large genus asymptotics of
$\langle \kappa^{3g-2-d}\psi_1^{d}\rangle_{g,1}$
using the technique of Richardson transformation.
We find
\begin{equation}
\begin{aligned}
 \frac{\langle \kappa^{3g-2-d}\psi_1^{d}\rangle_{g,1}}{(3g-2-d)!}
&=\frac{1}{\rt{8\pi}}\Bigl(\frac{2}{\pi^2}\Bigr)^{g}
\left(\frac{\pi}{2}\right)^{2d}\frac{\Ga(3/2)}{\Ga(3/2+d)}
\Biggl[\Ga(2g-3/2)\\
& +\left(-\frac{5}{24}
-\frac{6d^2-9d+11-6\cob_{d,0}}{3\pi^2}\right)\Ga(2g-3/2-1)+\cdots
\Biggr],\quad(g\gg d).
\end{aligned} 
\label{eq:asymp-int}
\end{equation}
For $d=0$ this agrees with the result in \cite{zograf1,zograf2}.
Plugging \eqref{eq:asymp-int} into the definition of $V_{g,1}(b)$
in \eqref{eq:Vg1}, we find the large genus asymptotics of $V_{g,1}(b)$ 
\begin{equation}
\begin{aligned}
 V_{g,1}(b)\sim \frac{2(4\pi^2)^{2g-3/2}}{(2\pi)^{3/2}}
 \sum_{n=0}^\infty f_n(b) \Ga(2g-3/2-n),
\quad (g\gg1)
\end{aligned} 
\label{eq:V-asym}
\end{equation}
with
\begin{equation}
\begin{aligned}
 f_0(b)&=\frac{2}{b}\sinh\frac{b}{2},\\
f_1(b)&=-\left(\frac{5}{24}+\frac{17}{3\pi^2}
+\frac{b^2}{8\pi^2}\right)\frac{2}{b}\sinh\frac{b}{2}+
\frac{2}{\pi^2}\left(1+
\cosh\frac{b}{2}\right).
\end{aligned} 
\end{equation}
$f_0(b)$ agrees with the result in \cite{Saad:2019lba}.
Note that the above $f_n(b)$ vanishes at $b=2\pi\ri$ which is
consistent with the property $V_{g,1}(2\pi\ri)=0$ \cite{zograf2}.
The large genus asymptotics in \eqref{eq:V-asym} implies that
there is a non-perturbative correction of the form
\begin{equation}
\begin{aligned}
 e^{-\frac{e^{S_0}}{4\pi^2}}=e^{-\frac{\pi}{2\hbar}}.
\end{aligned} 
\end{equation}
This is interpreted in \cite{Saad:2019lba} as the effect of ZZ brane 
\cite{Zamolodchikov:2001ah} sitting at $E=-\frac{\pi^2}{4}$ and the
instanton action agrees with the value of the effective potential
$V_{\text{eff}}(-\frac{\pi^2}{4})=\frac{\pi}{2}$.
(See \eqref{eq:V-eff} for the explicit form of $V_{\text{eff}}(E)$). 

%%%
\subsection{Genus-zero part of $Z_\JT$}
%%%

Let us revisit the genus-zero part of $Z_{\text{JT}}$
using our expression of the macroscopic loop operator
\eqref{eq:macro-loop}. 
At genus-zero, \eqref{eq:macro-loop} is reduced to
\begin{equation}
\begin{aligned}
  Z_{\text{JT}}^{(g=0)}
&=\int_{-\infty}^x dx' \int_{-\infty}^\infty \frac{dp}{2\pi}
e^{\bt(-p^2+ u_0(x'))}
=\frac{1}{2\rt{\pi\bt}}\int_{-\infty}^x dx' e^{\bt u_0(x')}.
\end{aligned} 
\label{eq:Zg0-int}
\end{equation} 
As discussed in \cite{Itzykson:1992ya},
the $x$-dependence of $u_0$ is determined from
the classical string equation \eqref{eq:classical-string}.
Recalling our definition $t_0=\hbar x$ in \eqref{eq:rescaledvar}, 
\eqref{eq:classical-string} is rewritten as
\begin{equation}
\begin{aligned}
 \hbar x=u_0-\sum_{k=1}^\infty t_k\frac{u_0^k}{k!}.
\end{aligned} 
\label{eq:g0-string-eq}
\end{equation}
For the on-shell value of the coupling $t_n=\ga_n~(n\geq1)$,
this becomes
\begin{equation}
\begin{aligned}
 \hbar x=u_0-\sum_{k=2}^\infty \frac{(-1)^k}{(k-1)!}\frac{u_0^k}{k!}
=\rt{u_0}J_1(2\rt{u_0}).
\end{aligned} 
\label{eq:x-vs-u}
\end{equation}
Then we can change the
integration variable in \eqref{eq:Zg0-int} from $x$ to $u_0$
via the relation \eqref{eq:x-vs-u}
\begin{equation}
\begin{aligned}
 Z_{\text{JT}}^{(g=0)}&
=\frac{1}{2\rt{\pi\bt}}
 \int_{-\infty}^{u_0} du\frac{\del x}{\del u} e^{\bt u}\\
&=\frac{1}{2\rt{\pi\bt}\hbar}
 \int_{-\infty}^{u_0}du J_0(2\rt{u})e^{\bt u}\\
&=\frac{1}{2\rt{\pi\bt}\hbar}
 \int^{\infty}_{-u_0}dv I_0(2\rt{v})e^{-\bt v}.
\end{aligned} 
\label{eq:ZJT-g0}
\end{equation}
Here $I_0(2\rt{v})$ denotes the modified Bessel function of
the first kind, which should not be confused with $I_0(u_0,\{t_k\})$.
Finally, using the relation
\begin{equation}
\begin{aligned}
 \int_v^\infty dE\frac{e^{-\bt E}}{\rt{E-v}}=\rt{\frac{\pi}{\bt}}e^{-\bt v},
\end{aligned} 
\end{equation}
we can recast $Z_{\text{JT}}^{(g=0)}$ in \eqref{eq:ZJT-g0} into the
integral of eigenvalue density
\begin{equation}
\begin{aligned}
 Z_{\text{JT}}^{(g=0)}&=\int_{-u_0}^\infty dE e^{-\bt E} \rho_0(E),\\
\rho_0(E)&=\int_{-u_0}^E\frac{dv}{2\pi\hbar}
\frac{I_0(2\rt{v})}{\rt{E-v}}.  
\end{aligned} 
\label{eq:rho-u0}
\end{equation}
When $u_0=0$ this reduces to the familiar form of
the eigenvalue density of the Schwarzian theory \cite{Stanford:2017thb}
\begin{equation}
\begin{aligned}
 \rho_0(E)&=\int_{0}^E\frac{dv}{2\pi\hbar}
\frac{I_0(2\rt{v})}{\rt{E-v}} =\frac{\sinh(2\rt{E})}{2\pi\hbar},
\end{aligned} 
\label{eq:rho0}
\end{equation}
and the genus-zero part of $Z_\JT$ in \eqref{eq:ZJT-g} is correctly
reproduced
\begin{equation}
\begin{aligned}
  Z_{\text{JT}}^{(g=0)}
&=\int_{0}^\infty dE e^{-\bt E}\frac{\sinh(2\rt{E})}{2\pi\hbar}
 =\frac{e^{\bt^{-1}}}{2\rt{\pi}\hbar\bt^{3/2}}.
\end{aligned} 
\end{equation}
To summarize, our expression of $Z_\JT$ as the macroscopic loop operator
$\Tr(e^{\bt Q}\Pi)$ in \eqref{eq:macro-Pi} automatically includes 
the contribution of disk topology $(g=0)$
as well as the higher genus $(g\geq1)$ corrections.
In other words, we do not have to treat the disk and other contributions
separately as in \eqref{eq:ZJT-g}. Our expression
$Z_\JT=\Tr(e^{\bt Q}\Pi)$ captures all contributions in one shot.

In the large $N$ limit of the matrix model, the genus-zero resolvent
obeys an algebraic equation which defines the so-called spectral curve.
In our normalization of $E$ and $\hbar$ in \eqref{eq:rho0},
the spectral curve of the matrix model is written as
\begin{equation}
\begin{aligned}
 y=\hf\sin(2\rt{\xi})=\hf\sin(2z),
\end{aligned} 
\label{eq:curve}
\end{equation}
where $E,\xi$ and $z$ are related by
\begin{equation}
\begin{aligned}
 E=-\xi=-z^2.
\end{aligned} 
\label{eq:E-xi}
\end{equation}
As utilized heavily in \cite{Saad:2019lba},
the genus expansion of matrix model is
essentially determined by the spectral curve via the
topological recursion \cite{Eynard:2007kz}.
However, to perform the actual computation of the genus expansion
with a fixed number of boundaries,
the topological recursion turns out to be a very slow algorithm
since to compute $V_{g,1}(b)$ we need to know all the data of $V_{g',n}$ 
with $g'+n\leq g+1~(g'\ge0, n\ge1)$.
As emphasized in \cite{zograf1},
the method of KdV equation in section \ref{sec:genus}
provides us with a very fast algorithm for the computation of the genus
expansion at a fixed number of boundaries.

%%%
\subsection{Low temperature expansion}\label{sec:low}
%%%

As we saw in section~\ref{sec:genus}, 
the intersection number $\bra \ka^\ell \psi_1^{3g-2-\ell}\ket_g$ with 
small $\ell$ can be computed for all genus.
We observed that
the values for fixed $\ell$ are
governed by the polynomial $P_\ell(g)$.
Based on this observation we expect that
one can write the expansion of
$\langle Z(\beta)\rangle=Z_{\text{JT}}(0,0)$ as
\begin{equation}
\label{eq:Z00Texp}
\begin{aligned}
\langle Z(\beta)\rangle
&=\frac{1}{\rt{2\pi}\gs\bt^{3/2}}
 \left[e^{T}+\sum_{g=1}^\infty (\gs\bt^{3/2})^{2g}
 \sum_{\ell=0}^{3g-2}\frac{T^\ell}{\ell!}
 \frac{P_{\ell}(g)}{24^g g!}\right]\\
&=\frac{1}{\rt{2\pi}\gs\bt^{3/2}}
 \sum_{\ell=0}^\infty\frac{T^\ell}{\ell!}
 \left[1+\sum_{g=1}^\infty
  (\gs\bt^{3/2})^{2g}\frac{P_{\ell}(g)}{24^g g!}\right]\\
&=\frac{1}{\rt{2\pi}\gs\bt^{3/2}}
 \sum_{\ell=0}^\infty\frac{T^\ell}{\ell!}
 \sum_{g=0}^\infty(\gs\bt^{3/2})^{2g}\frac{P_{\ell}(g)}{24^g g!}.
\end{aligned} 
\end{equation}
When going from the first line to the second line of \eqref{eq:Z00Texp}
we have removed the restriction
$3g-2\geq\ell$ since
$P_\ell(g)=(24)^gg!\langle\kappa^\ell\psi_1^{3g-2-\ell}\rangle_{g,1}$
vanishes when $3g-2<\ell$.
In the last equality of \eqref{eq:Z00Texp}
we used the property $P_{\ell}(0)=1$ to extend the summation
to $g=0$. 
It is natural to expect that
$Z_\JT(t_0,t_1)$ also admits a similar low temperature expansion,
namely a power series expansion in $T$ with $\gs\beta^{3/2}$ being
fixed. In what follows we will see that such an expansion
indeed exists and can be computed
with the help of the differential equation \eqref{eq:Wdiffeq}
and the results of the genus expansion of $u$.

To begin with, 
let us consider the all-genus resummation of the $\ell=0$ term
in \eqref{eq:Z00Texp}
\begin{equation}
\begin{aligned}
 \frac{1}{\rt{2\pi}\gs\bt^{3/2}}
\sum_{g=0}^\infty(\gs\bt^{3/2})^{2g}\frac{1}{24^g g!}
 =\frac{e^{\frac{\gs^2\bt^3}{24}}}{\rt{2\pi}\gs\bt^{3/2}}.
\end{aligned} 
\end{equation}
We notice that this is essentially the partition function
in the Airy case
\begin{equation}
\begin{aligned}
Z_{\text{Airy}}
 =\int_{-\infty}^\infty dE \rho_{\text{Airy}}(E)e^{-\bt E}
 =\frac{e^{\frac{\hbar^2\bt^3}{12}}}{2\rt{\pi}\hbar\bt^{3/2}},
\end{aligned} 
\label{eq:z-airy}
\end{equation}
where
\begin{equation}
\begin{aligned}
 \rho_{\text{Airy}}(E)=\hbar^{-\frac{2}{3}}
\left[\text{Ai}'(-\hbar^{-\frac{2}{3}}E)^2
  -\text{Ai}(-\hbar^{-\frac{2}{3}}E)
   \text{Ai}''(-\hbar^{-\frac{2}{3}}E)\right]
\end{aligned} 
\label{eq:rho-airy}
\end{equation}
and $\hbar$ is related to $\gs$ by \eqref{eq:rescaledvar}.
(See appendix \ref{app:airy} for the summary of the Airy case).
In what follows we will use $\hbar$ as the genus counting parameter
instead of $\gs$.

One can generalize the exponential factor in \eqref{eq:z-airy} 
to the off-shell value, as follows.
When computing $Z_\JT$ from $W=\partial_x Z_\JT$,
the above $\ell=0$ term at each genus is originated from
the $\mathcal{O}(\bt^{3g})$ term in $W_g$.
We observe that the $\mathcal{O}(\bt^{3g})$ term,
which is of highest order in $\beta$, appears in $W_g$ as
\begin{equation}
\begin{aligned}
 W_g=\frac{\bt^{3g}}{24^gg!t^{2g}}+\cdots.
\end{aligned}
\end{equation}
This gives rise to the factor
\begin{equation}
\begin{aligned}
 \sum_{g=0}^\infty (\rt{2}\hbar)^{2g}\frac{\bt^{3g}}{24^gg!t^{2g}}
=e^{\frac{h^2}{12t^2}},
\end{aligned} 
\label{eq:h-factor}
\end{equation}
where we have defined
\begin{equation}
\begin{aligned}
 h:=\hbar\bt^{3/2}.
\end{aligned} 
\label{eq:def-h}
\end{equation}
Thus it is natural to make an ansatz
\begin{equation}
\begin{aligned}
W=\frac{\sqrt{T}}{2\sqrt{\pi}}e^{\frac{h^2}{12t^2}+\frac{y}{T}}W_L,\quad
W_L=\sum_{\ell=0}^\infty T^\ell w_\ell(h),
\end{aligned} 
\label{eq:WLexp}
\end{equation}
where we have factored out the genus-zero part \eqref{eq:Wg=0} and the
exponential in \eqref{eq:h-factor} as the prefactor,
so that we have $w_0=1$.
In the rest of this section
we regard $T$ and $h$ as the independent parameters.
Plugging this ansatz into \eqref{eq:W-eq}, we find
\begin{equation}
\label{eq:WLdiffeq}
\begin{aligned}
 -\del_t W_L+\frac{h^2}{6t^3}W_L=\h{u}DW_L+\frac{h^2T^3}{6}D^3W_L,
\end{aligned} 
\end{equation}
where $D$ is given by
\begin{equation}
\begin{aligned}
D=e^{-\frac{h^2}{12t^2}
 -\frac{y}{T}}\del_0 e^{\frac{h^2}{12t^2}+\frac{y}{T}}
 =\del_0+\frac{1}{tT}+\frac{h^2 B_1}{6t^4}
\end{aligned} 
\end{equation}
and $\h{u}=u-y$ is given by 
\begin{equation}
\begin{aligned}
 \h{u}=\sum_{g=1}^\infty\gs^{2g}u_g
 =\sum_{g=1}^\infty (\rt{2}h)^{2g}T^{3g}u_g.
\end{aligned} 
\end{equation}
Here $u_g$ is determined by the recursion relation \eqref{eq:urecrel}.
By plugging the expansion \eqref{eq:WLexp}
into the differential equation \eqref{eq:WLdiffeq}
and integrating the ${\cal O}(T^\ell)$ part with respect to $t$,
one can recursively compute $w_\ell$
starting with the initial condition $w_0=1$.
To determine $w_\ell$ uniquely one needs to require that
$w_\ell\ (\ell\ge 1)$ does not contain
any ${\cal O}(t^0)$ term.
This can be shown as follows:
since $W_g\ (g\ge 1)$ is a polynomial in $t^{-1}$
without any ${\cal O}(t^0)$ term,
$W_L-1=\exp(-\frac{h^2}{12}t^{-2})\sum_{g=0}^\infty \gs^{2g}W_g-1$
is a formal power series in $t^{-1}$ without any ${\cal O}(t^0)$ term
and so does $w_\ell\ (\ell\ge 1)$.
Consequently,
\eqref{eq:WLdiffeq} unambiguously determines $w_\ell$.
The first two terms of $w_\ell$ are given by
\begin{equation}
\begin{aligned}
w_1&=\left(\frac{h^2}{6t^3}+\frac{h^4}{60t^5}\right)B_1,\\
w_2&=
  \left(\frac{h^2}{6t^4}+\frac{h^4}{8t^6}+\frac{7h^6}{720t^8}
    +\frac{h^8}{7200t^{10}}\right)B_1^2
 -\left(\frac{h^2}{12t^3}+\frac{h^4}{30t^5}
    +\frac{h^6}{840t^7}\right)B_2.
\end{aligned} 
\label{eq:well-first}
\end{equation}
It turns out that $w_\ell$ is actually a polynomial
of weight $(\ell,0)$
in the generators $B_n\ (n\ge 1)$, $t^{-1}$ and $h$,
to which weights $(n,1)$, $(0,-1)$ and $(0,1)$
are assigned respectively.

Finally, let us expand $Z_\JT$ as
\begin{equation}
\begin{aligned}
Z_\JT=\frac{e^{\frac{h^2}{12t^2}+\frac{y}{T}}}{2\sqrt{\pi}h}Z_L,\quad
Z_L=\sum_{\ell=0}^\infty \frac{T^\ell}{\ell!} z_\ell.
\end{aligned} 
\label{eq:ZL-ansatz}
\end{equation}
The relation $\partial_x Z_\JT=W$ is rewritten as
\begin{equation}
\begin{aligned}
TDZ_L=W_L.
\end{aligned}
\label{eq:ZL-WL} 
\end{equation}
Comparing the coefficient of $T^{\ell}$ on both sides
of \eqref{eq:ZL-WL} we find
the recursion relation for $z_\ell~(\ell\ge0)$
\begin{equation}
\begin{aligned}
 z_\ell
 =t\left[\ell!w_{\ell}
 -\ell\left(\del_0+\frac{h^2B_1}{6t^4}\right)z_{\ell-1}\right],
\end{aligned} 
\end{equation}
where we formally set $z_{-1}=0$.
For instance, using $w_0=1$
and the result of $w_{1,2}$ in \eqref{eq:well-first}
we find
\begin{equation}
\begin{aligned}
z_0&=t,\\
z_1&=\left(1+\frac{h^4}{60t^4}\right)B_1,\\
z_2&=\left(\frac{7h^4}{60t^5}+\frac{h^6}{72t^7}
   +\frac{h^8}{3600t^9}\right)B_1^2
 +\left(2-\frac{h^2}{6t^2}-\frac{h^4}{30t^4}
   -\frac{h^6}{420t^6}\right)B_2.
\end{aligned} 
\end{equation}

For the on-shell value $(y,t)=(0,1)$, \eqref{eq:ZL-ansatz} becomes
\begin{equation}
\begin{aligned}
 Z_\JT=\frac{e^{\frac{h^2}{12}}}{2\rt{\pi}h}
  \sum_{\ell=0}^\infty\frac{T^\ell}{\ell!}\tz_\ell(h),
\end{aligned} 
\label{eq:Z-tz}
\end{equation}
where $\tz_\ell := z_\ell\big|_{y=0,t=1}$.
Note that $\tz_\ell(h)$ can be thought of as the generating function for
the intersection numbers
$\langle \kappa^\ell\psi_1^{3g-2-\ell}\rangle_{g,1}$
\begin{equation}
\begin{aligned}
 \sum_{g=0}^\infty({\rt{2}h})^{2g}
  \langle \kappa^\ell\psi_1^{3g-2-\ell}\rangle_{g,1}
=\sum_{g=0}^\infty h^{2g}\frac{P_\ell(g)}{(12)^gg!}
=e^{\frac{h^2}{12}}\tz_\ell(h).
\end{aligned} 
\end{equation}
The first few of them read
\begin{equation}
\begin{aligned}
\tz_1(h)&=1+\frac{h^4}{60},\\
\tz_2(h)&=1-\frac{h^2}{12}+\frac{h^4}{10}+\frac{4h^6}{315}
  +\frac{h^8}{3600},\\
\tz_3(h)&=1-\frac{h^2}{12}+\frac{16h^4}{45}+\frac{1163h^6}{5040}
  +\frac{13h^8}{560}+\frac{h^{10}}{1575}+\frac{h^{12}}{216000}.
\end{aligned} 
\label{eq:zh}
\end{equation}
We have computed $\tz_\ell(h)$ up to $\ell=50$.\footnote{
The data of $\tz_\ell(h)~(\ell=1,\cdots,50)$ are attached to
the arXiv submission in the file \texttt{zlowdata.txt}.
}
The method above serves as another very efficient algorithm
for computing the intersection number
$\langle \kappa^\ell\psi_1^{3g-2-\ell}\rangle_{g,1}$,
in particular for large $g$ with fixed $\ell$.

Note that $\tz_\ell(h)$ is obtained from $P_\ell(g)$ as 
\begin{align}
\begin{aligned}
\tz_\ell(h)&=e^{-\frac{h^2}{12}}
 P_\ell\left(\frac{h}{2}\frac{\partial}{\partial h}\right)
 e^{\frac{h^2}{12}}
=P_\ell
 \left(\frac{h^2}{12}+\frac{h}{2}\frac{\partial}{\partial h}\right)
 \cdot 1.
\end{aligned}
\end{align}
Conversely, noticing that
\begin{align}
 \left(\frac{h^2}{12}+\frac{h}{2}\frac{\partial}{\partial h}\right)^k
 \cdot 1
=\frac{h^{2k}}{12^k}+\mbox{lower order terms in $h$},
\end{align}
one can easily calculate $P_\ell(g)$ from $\tz_\ell(h)$
by iteratively determining the coefficients of
$g^k\ (0\le k\le 2\ell)$ in descending order.
We have thus obtained $P_\ell(g)$ also up to $\ell=50$.
As we explained below \eqref{eq:Z00Texp},
$P_\ell(g)$ vanishes for $\{g\in\bbZ_{> 0}\,|\,3g-2<\ell\}$.
We have verified that our results indeed satisfy this property.

%%%%%%%%%%%%%%%%%%%%%%%%%%%%%%%%%%%%%%%%%%%%%%%%%%%%%%%%%%%%%%%%%%%%%%%%
\section{Various limits in the low temperature regime}\label{sec:limit}
%%%%%%%%%%%%%%%%%%%%%%%%%%%%%%%%%%%%%%%%%%%%%%%%%%%%%%%%%%%%%%%%%%%%%%%%

In the low temperature regime we can take various limits of $Z_\JT$ 
and the BA function $\psi(E)$.
In this section we will consider the low energy limit of $\rho(E)$
and $\psi(E)$,
and also the 't Hooft limit of $Z_\JT$ and the Laplace transform $\h{\psi}$ of the BA function.
In the rest of this paper we will turn off the deformation parameter 
$t_0=t_1=0$ and consider the
partition function $Z_\JT$ and related quantities at the on-shell value of $t_n=\ga_n~(n\geq0)$.

%%%
\subsection{Low energy expansion of $\rho(E)$}
\label{sec:low-rho}
%%%
In this subsection we will consider the low energy expansion of the eigenvalue density
$\rho(E)$
in the limit
\begin{equation}
\begin{aligned}
 \hbar,E\to0\quad \text{with}~~\eta=-\hbar^{-\frac{2}{3}}E~~\text{fixed}.
\end{aligned} 
\label{eq:small-E-lim}
\end{equation}
In this limit the genus-zero part $\rho_0(E)$ in \eqref{eq:rho0} is expanded as
\begin{equation}
\begin{aligned}
 \rho_0(E)dE&=\frac{dE}{\pi\hbar}
\sum_{\ell=0}^\infty\frac{2^{2\ell}E^{\ell+\hf}}{(2\ell+1)!}
=-\frac{d\eta}{\pi}\sum_{\ell=0}^\infty \hbar^{\frac{2}{3}\ell}\frac{2^{2\ell}(-\eta)^{\ell+\hf}}
{(2\ell+1)!}.
\end{aligned} 
\label{eq:small-E-rho}
\end{equation}
Let us consider the first term of this expansion
\begin{equation}
\begin{aligned}
 \rho_0(E)dE&=\frac{\sinh(2\rt{E})dE}{2\pi\hbar}=\frac{\rt{E}dE}{\pi\hbar}+\cdots.
\end{aligned} 
\label{eq:rho0-expand}
\end{equation}
It is well-known that this term corresponds to the Airy case as 
reviewed in appendix \ref{app:airy}.
In this case the genus-zero eigenvalue density in \eqref{eq:rho0-expand}
is promoted to the full eigenvalue density $\rho_{\text{Airy}}(E)$ in \eqref{eq:rho-airy}.
It turns out that each term of this expansion
\eqref{eq:small-E-rho}
has its own all-genus completion
\begin{equation}
\begin{aligned}
 \rho(E)dE=-\sum_{\ell=0}^\infty \hbar^{\frac{2}{3}\ell} \varrho_\ell(\eta)d\eta,
\end{aligned} 
\label{eq:rho-eta}
\end{equation}
where $\varrho_\ell(\eta)$ is defined in such a way that it 
reduces to the $\ell$-th term in the expansion 
of $\rho_0(E)$ in \eqref{eq:small-E-rho}
in the classically allowed region $E>0$
\begin{equation}
\begin{aligned}
 \lim_{\eta\to-\infty}\varrho_\ell(\eta)=\frac{2^{2\ell}(-\eta)^{\ell+\hf}}{\pi (2\ell+1)!},
\end{aligned} 
\label{eq:rho-lim}
\end{equation}
up to an oscillatory correction.
As we discussed above, $\varrho_0(\eta)$ is given by $\rho_{\text{Airy}}(E)$ 
in \eqref{eq:rho-airy} up to a normalization factor
\begin{equation}
\begin{aligned}
 \varrho_0(\eta)=\text{Ai}'(\eta)^2-\eta \text{Ai}(\eta)^2.
\end{aligned} 
\label{eq:varrho-airy}
\end{equation}
In terms of the coupling $h$ defined in \eqref{eq:def-h}, 
$Z_{\text{Airy}}$ in \eqref{eq:z-airy} is written as
\begin{equation}
\begin{aligned}
\int_{-\infty}^\infty d\eta \,e^{-\bt E}\varrho_0(\eta)=\int_{-\infty}^\infty d\eta \,e^{h^{\frac{2}{3}}\eta}\varrho_0(\eta)= \frac{e^{\frac{h^2}{12}}}{2\rt{\pi}h}.
\end{aligned}
\label{eq:varrho0-int} 
\end{equation}

We can determine the higher order terms $\varrho_\ell(\eta)$ by matching
the low temperature expansion of $Z_\JT$ in \eqref{eq:Z-tz}
\begin{equation}
\begin{aligned}
 Z_\JT&=\frac{e^{\frac{h^2}{12}}}{2\rt{\pi}h}\sum_{\ell=0}^\infty \frac{T^\ell}{\ell!} \tz_\ell(h)
=\frac{e^{\frac{h^2}{12}}}{2\rt{\pi}h}\sum_{\ell=0}^\infty \frac{1}{\ell!}\left(\frac{\hbar}{h}\right)^{\frac{2}{3}\ell} \tz_\ell(h).
\end{aligned} 
\label{eq:Z-low}
\end{equation}
From the definition of eigenvalue density
\begin{equation}
\begin{aligned}
 Z_\JT=\int_{-\infty}^\infty dE\rho(E)e^{-\bt E},
\end{aligned} 
\end{equation}
the expansion of $Z_\JT$ \eqref{eq:Z-low} 
and the expansion of $\rho(E)$ \eqref{eq:rho-eta} become consistent if $\varrho_\ell(\eta)$
satisfies
\begin{equation}
\begin{aligned}
 \int_{-\infty}^\infty d\eta \,e^{-\bt E}\varrho_\ell(\eta)= 
\int_{-\infty}^\infty d\eta \,e^{h^{\frac{2}{3}}\eta}\varrho_\ell(\eta)
=\frac{e^{\frac{h^2}{12}}}{2\ell!\rt{\pi}h^{1+\frac{2}{3}\ell}}\tz_\ell(h).
\end{aligned} 
\label{eq:rho-ell-z}
\end{equation}
Using \eqref{eq:varrho0-int}, this relation  can be formally solved as
\begin{equation}
\begin{aligned}
 \varrho_\ell(\eta)=\frac{\tz_\ell\big((-\del_\eta)^{3/2}\big)}{\ell!(-\del_\eta)^\ell}\varrho_0(\eta).
\end{aligned} 
\label{eq:formal-rho}
\end{equation}
For instance, from the result of $\tz_1(h)$ in \eqref{eq:zh}, $\varrho_1(\eta)$ is given by
\begin{equation}
\begin{aligned}
 \varrho_1(\eta)&=-\del_\eta^{-1}\varrho_0(\eta)-\frac{\del_\eta^5}{60}\varrho_0(\eta)
=\frac{2}{15}\Bigl[6\eta^2\text{Ai}(\eta)^2-\text{Ai}(\eta)\text{Ai}'(\eta)-4\eta 
\text{Ai}'(\eta)^2\Bigr].
\end{aligned} 
\end{equation}
Here the negative power of $\del_\eta$ should be understood as
the integration with respect to $\eta$. 
One might think that there is an ambiguity in the integration constant,
but $\varrho_\ell(\eta)$ is actually determined unambiguously by requiring \eqref{eq:rho-ell-z}.
Using the data of $\tz_\ell(h)$ in \eqref{eq:zh} we find
\begin{equation}
\begin{aligned}
  \varrho_2(\eta)&=\frac{152 \eta^2}{1575}\text{Ai}'(\eta)^2
-\left(\frac{296
   \eta^3}{1575}+\frac{3}{200}\right)\text{Ai}(\eta)^2
-\left(\frac{16 \eta^4}{225}+\frac{8
   \eta}{525}\right)\text{Ai}(\eta)\text{Ai}'(\eta),\\
 \varrho_3(\eta)&=\left(\frac{64 \eta^7}{10125}+\frac{16
   \eta^4}{567}-\frac{37 \eta}{8100}
\right)\text{Ai}(\eta)^2
+\left(\frac{704
   \eta^5}{23625}+\frac{\eta^2}{189}
\right)\text{Ai}(\eta)\text{Ai}'(\eta)\\
&\hskip10mm+\left(
\frac{64 \eta^6}{10125}-\frac{32
   \eta^3}{14175}-\frac{19}{8100}
\right)\text{Ai}'(\eta)^2.
\end{aligned} 
\end{equation}
This procedure enables us to find the all-genus completion of 
the eigenvalue density order by order in
the small $E$ expansion \eqref{eq:small-E-rho}.
In general, $\varrho_\ell(\eta)$ is written as a combination of the Airy function $\text{Ai}(\eta)$
and its derivatives.
This implies that $\varrho_\ell(\eta)$ is exponentially small
in the classically forbidden region $\eta>0$, which is 
indeed necessary for the convergence of the integral \eqref{eq:rho-ell-z}.
In appendix \ref{app:partial}, we consider a partial resummation
of this expansion of $\rho(E)$.

%%%
\subsection{'t Hooft expansion of $Z_\JT$}
\label{sec:thooft-Z}
%%%
In the low temperature regime
we can take the 't Hooft limit \eqref{eq:tHooft}.
As we will see shortly, the relation between $Z_\JT$ and the spectral curve
becomes manifest in this limit.

We can rearrange the low temperature expansion 
in terms of the parameters $\la$ and $\hbar$ in \eqref{eq:tHooft}.
Plugging the relation
\begin{equation}
\begin{aligned}
 h=\left(\frac{\la^3}{\hbar}\right)^\hf,\quad
T=\frac{\hbar}{\la}
\end{aligned} 
\end{equation}
into the low temperature expansion of $Z_\JT$
in \eqref{eq:Z-low}, we find that the free energy
is expanded as \eqref{eq:thooft-free}.
From the data of $\tz_\ell(h)$ obtained in the previous section, we can compute
$\cF_n(\la)$ in \eqref{eq:thooft-free} as a power series expansion in $\la$.
By matching the first few orders of this series expansion, we find the closed form 
of $\cF_{0}(\la)$
\begin{equation}
\begin{aligned}
 \cF_0(\la)
&=\qu\la\arcsin(\la)^2+\hf\Bigl(
\rt{1-\la^2}\arcsin(\la)-\lambda\Bigr).
\end{aligned} 
\label{eq:CF-0}
\end{equation}
One can show that this is written as
\begin{equation}
\begin{aligned}
 \cF_0(\la)=2\int^{\la/2}_0 \xi(y)dy,
\end{aligned} 
\label{eq:F0-period}
\end{equation}
where $\xi(y)$ is determined by the spectral curve \eqref{eq:curve}
\begin{equation}
\begin{aligned}
 y=\hf\sin(2\rt{\xi})\quad\Rightarrow\quad\xi(y)=\qu\arcsin(2y)^2.
\end{aligned} 
\end{equation}
Namely, $\cF_0(\la)$ is given by the integral of one-form
$\xi dy$ on the spectral curve.
Recall that the effective potential $V_{\text{eff}}(E)$
for the eigenvalue is given by the integral of another one-form $yd\xi$
\begin{equation}
\begin{aligned}
 V_{\text{eff}}(E)=2\int_0^{-E}yd\xi=\hf \sin(2\rt{-E})-\rt{-E}\cos(2\rt{-E}).
\end{aligned} 
\label{eq:V-eff}
\end{equation}
As we will discuss in section \ref{sec:WKB},
the appearance of the ``dual'' one-form $\xi dy$ in \eqref{eq:F0-period}
can be  understood
from the Laplace transformation.

From the data of series expansion, we also find the closed form of $\cF_{1,2}(\la)$
\begin{equation}
\begin{aligned}
\cF_1(\la)&=-\frac{3}{2}\log\arcsin(\la)-\qu\log(1-\la^2)+\hf\log\frac{\hbar}{4\pi},\\
\cF_2(\la)&=\frac{17}{3\arcsin(\la)^3}\left[-1+\frac{1}{\rt{1-\la^2}}\right]
-\frac{23\la}{12(1-\la^2)\arcsin(\la)^2}\\
&+\frac{1}{12\arcsin(\la)}\left[-2-\frac{2}{\rt{1-\la^2}}+\frac{5}{(1-\la^2)^{3/2}}\right].
\end{aligned} 
\label{eq:CF-n}
\end{equation}
In section \ref{sec:WKB}, we will see that $\cF_2(\la)$
can be obtained analytically from the result of topological recursion.
Apparently, the above form of $\cF_n(\la)$ becomes singular at $\la=1$, and 
\eqref{eq:CF-n} can be trusted only in the region $\la<1$.
If we analytically continue $\cF_n(\la)$ to complex $\la$, 
there is a cut running from $\la=1$ to $\la=+\infty$ along the real axis
of complex $\la$-plane.
It is interesting to understand the physical origin of the
singularity at $\la=1$.

Before closing this section, we comment on the genus expansion of free energy 
$\cF=\log Z_\JT$.
In the original parameters $(\gs,\bt)$ without taking any particular limit, the free energy
admits the ordinary genus expansion
\begin{equation}
\begin{aligned}
 \cF&=\sum_{g=0}^\infty \gs^{2g}\til{\cF}_g(\bt)
=\frac{1}{\bt}-\log\big(\rt{2\pi} \gs\bt^{3/2}\big)+\gs^2\frac{\bt^3+\bt^2}{24}e^{-\frac{1}{\bt}}+\mathcal{O}(\gs^4).
\end{aligned} 
\label{eq:high-free}
\end{equation}
This expansion is valid in the high temperature regime $\bt\ll1$.
On the other hand, in the low temperature regime in the 't Hooft limit
\eqref{eq:tHooft}, the free energy
is expanded as \eqref{eq:thooft-free}.
One can recognize that
the high temperature expansion \eqref{eq:high-free} is ``closed string'' like, while
the low temperature expansion \eqref{eq:thooft-free} is ``open string'' like.
$\gs$ and $\hbar$ can be thought of as the closed string coupling and the
open string coupling, respectively.

%%%
\subsection{Low energy expansion of $\psi(E)$}
%%%
In this section we will consider the low energy expansion of BA function
$\psi(E)$
in the limit \eqref{eq:small-E-lim}.
This expansion is easily obtained from
the expansion of $W=\bra x|e^{\bt Q}|x\ket$ by using the relation
\begin{equation}
\begin{aligned}
 W=\int_{-\infty}^\infty dE\bra x|E\ket e^{-\bt E}\bra E|x\ket=
\hbar\int_{-\infty}^\infty dE\, e^{-\bt E}\psi(E)^2.
\end{aligned} 
\end{equation}
Here we have put the extra factor of $\hbar$ to match the
result of Airy case in \eqref{eq:BA-airy}. 
From the low temperature expansion of $W$
\begin{equation}
\begin{aligned}
 W=\frac{e^{\frac{h^2}{12}}}{2\rt{\pi}}\sum_{\ell=0}^\infty T^{\ell+\hf}w_\ell(h)
=\frac{e^{\frac{h^2}{12}}}{2\rt{\pi}h^{\frac{1}{3}}}\sum_{\ell=0}^\infty
\hbar^{\frac{2\ell+1}{3}}\frac{w_\ell(h)}{h^{\frac{2\ell}{3}}},
\end{aligned} 
\end{equation}
we can compute the low energy expansion of $\psi(E)^2$ starting from the relation
\begin{equation}
\begin{aligned}
 \int_{-\infty}^\infty d\eta \,e^{h^{\frac{2}{3}}\eta}\text{Ai}(\eta)^2=
\frac{e^{\frac{h^2}{12}}}{2\rt{\pi}h^{\frac{1}{3}}}.
\end{aligned} 
\end{equation}
As in the case of $\varrho_\ell(\eta)$ in \eqref{eq:formal-rho},
$\psi(E)^2$ can be formally written as
\begin{equation}
\begin{aligned}
 \psi(E)^2=\sum_{\ell=0}^\infty \hbar^{\frac{2\ell-4}{3}}
\frac{w_\ell\bigl((-\del_\eta)^{3/2}\bigr)}{(-\del_\eta)^\ell}\text{Ai}(\eta)^2.
\end{aligned} 
\end{equation}
From this expansion we can easily find the  expansion
of $\psi(E)$ in the low energy limit \eqref{eq:small-E-lim}
\begin{equation}
\begin{aligned}
 \psi(E)=
\sum_{\ell=0}^\infty \hbar^{\frac{2}{3}(\ell-1)}\Psi_\ell(\del_\eta)\text{Ai}(\eta).
\end{aligned} 
\label{eq:psi-low}
\end{equation}
The first few terms of the differential operators $\Psi_\ell$ read
\begin{equation}
\begin{aligned}
\Psi_0&=1,\\
\Psi_1&=-\frac{4 \del_\eta^5}{15}+\del_\eta^2,\\
\Psi_2&=\frac{8 \del_\eta^{10}}{225}-\frac{212
   \del_\eta^7}{315}+\frac{5
   \del_\eta^4}{2}-\frac{9 \del_\eta}{8},\\
\Psi_3&=-\frac{32
   \del_\eta^{15}}{10125}+\frac{136
   \del_\eta^{12}}{945}-\frac{66
   \del_\eta^9}{35}+\frac{39
   \del_\eta^6}{5}-\frac{1655
   \del_\eta^3}{216}+\frac{11}{24}.
\end{aligned} 
\label{eq:Psi}
\end{equation}

In a similar manner as in section \ref{sec:thooft-Z}, 
we can consider the 't Hooft limit of the Laplace transform of $\psi(E)$.
Plugging the integral representation of Airy function
\begin{equation}
\begin{aligned}
 \text{Ai}(\eta)=\int_C\frac{dv}{2\pi\ri}e^{\frac{v^3}{3}-\eta v}=\int_C
\frac{d\la}{4\pi\ri\hbar^{\frac{1}{3}}}e^{\frac{\la^3}{24\hbar}+\frac{E\la}{2\hbar}}
\end{aligned} 
\label{eq:airy-C}
\end{equation}
into the expansion of $\psi(E)$ in \eqref{eq:psi-low}, we find
the 't Hooft expansion of the Laplace transform $\h{\psi}(\la)$ of the BA function $\psi(E)$ 
\begin{equation}
\begin{aligned}
 \psi(E)&=\int_C
\frac{d\la}{4\pi\ri\hbar}e^{\frac{\la^3}{24\hbar}+\frac{E\la}{2\hbar}}\sum_{\ell=0}^\infty \hbar^{\frac{2}{3}\ell}\Psi_\ell\Bigl(-\hf \la\hbar^{-\frac{1}{3}}\Bigr)
=:\int_C
\frac{d\la}{4\pi\ri\hbar}e^{\frac{E\la}{2\hbar}}\h{\psi}(\la).
\end{aligned}  
\label{eq:psi-Lap}
\end{equation}
More explicitly, $\h{\psi}(\la)$ is written as
\begin{equation}
\begin{aligned}
 \h{\psi}(\la)=e^{\frac{\la^3}{24\hbar}}\sum_{\ell=0}^\infty \hbar^{\frac{2}{3}\ell}\Psi_\ell\Bigl(-\hf \la\hbar^{-\frac{1}{3}}\Bigr).
\end{aligned} 
\end{equation}
In \eqref{eq:airy-C} the integration contour $C$ is chosen as the so-called Airy contour running
from $e^{-\frac{\pi\ri}{3}}\infty$ to $e^{\frac{\pi\ri}{3}}\infty$ on the complex $\la$-plane.
As in the case of the partition function $Z_\JT$, $\h{\psi}(\la)$ admits 
the open string like expansion
\begin{equation}
\begin{aligned}
 \h{\psi}(\la)=\exp\Biggl(\sum_{n=0}^\infty \hbar^{n-1}G_n(\la)\Biggr).
\end{aligned} 
\label{eq:hat-psi}
\end{equation}
From the data of $\Psi_\ell$ in \eqref{eq:Psi}, we can compute $G_n(\la)$ as a series expansion
in $\la$. By matching the first few orders of the series expansion, 
we find the closed form of $G_n(\la)$
\begin{equation}
\begin{aligned}
 G_0(\la)&=\hf \cF_0(\la), \\
G_1(\la)&=-\qu\log(1-\la^2),\\
G_2(\la)&=\frac{5}{6\arcsin(\la)^3}\left[-1+\frac{1}{\rt{1-\la^2}}\right]+\frac{5\la}{6(1-\la^2)
\arcsin(\la)^2}\\
&\quad +\frac{1}{\arcsin(\la)}\left[\frac{5}{6(1-\la^2)^{3/2}}-\frac{1}{3\rt{1-\la^2}}-\frac{1}{12}\right],
\end{aligned} 
\label{eq:Gn-result}
\end{equation}
where $\cF_0(\la)$ is given by \eqref{eq:CF-0}.
Again, in the next section we will see that
$G_2(\la)$ can be obtained analytically from the topological recursion. 

%%%
\subsection{WKB expansion of $\psi(E)$ and $\rho(E)$ from topological recursion}
\label{sec:WKB}
%%%
In this section we will systematically compute the semi-classical $\hbar$-expansion
(WKB expansion) of $\psi(E)$ and $\rho(E)$ from the topological recursion.

\subsubsection{WKB expansion of $\psi(E)$}\label{sec:wkb-psi}
Let us first consider the WKB expansion of $\psi(E)$.
Once we know the WKB expansion of $\psi(E)$,
the expansion of $\h{\psi}(\la)$ can be obtained from the saddle point approximation
of the integral
\begin{equation}
\begin{aligned}
 \h{\psi}(\la)=\int_{-\infty}^\infty dE \,e^{-\frac{E \la}{2\hbar}}\psi(E)
=\int_{-\infty}^\infty d\xi \,e^{\frac{\xi\la}{2\hbar}}\psi(-\xi),
\end{aligned} 
\label{eq:BA-lap}
\end{equation}
where $E$ and $\xi$ are related by \eqref{eq:E-xi}.
The BA function has the following WKB expansion
\begin{equation}
\begin{aligned}
 \psi(E)=\exp\Biggl(\sum_{n=0}^\infty \hbar^{n-1}S_n(\xi)\Biggr).
\end{aligned} 
\end{equation}
It is well-known that the leading term $S_0(\xi)$ is given by
the integral of one-form $yd\xi$
on the spectral curve \eqref{eq:curve} (see appendix \ref{app:BA} for a review)
\begin{equation}
\begin{aligned}
 S_0(\xi)=-\int_0^\xi y(\xi')d\xi'= -\hf V_{\text{eff}}(-\xi),
\end{aligned} 
\end{equation}
where $V_{\text{eff}}$ is given by \eqref{eq:V-eff}.
In the limit $\hbar\to0$, we can evaluate the
integral \eqref{eq:BA-lap} by the saddle point approximation.
The saddle point $\xi_*$ of \eqref{eq:BA-lap} is given by
\begin{equation}
\begin{aligned}
 \la-\sin(2\rt{\xi_*})=0\quad\Rightarrow\quad \xi_*=\qu\arcsin(\la)^2.
\end{aligned} 
\end{equation}
Then the leading term $G_0(\la)$ in the $\hbar$-expansion of $\h{\psi}(\la)$
in \eqref{eq:hat-psi} becomes
\begin{equation}
\begin{aligned}
 G_0(\la)=\frac{\xi_*\la}{2}+S_0(\xi_*)=
\int_0^{\la/2}\xi(y)dy=\hf\cF_0(\la).
\end{aligned} 
\end{equation}
As advertised, the integral of dual one-form $\xi dy$ naturally arises from
the saddle point approximation of \eqref{eq:BA-lap}.

Let us proceed to the higher order corrections.
Using the fact that the BA function
is the expectation value of the determinant operator
\begin{equation}
\begin{aligned}
 \psi(E)=e^{-\frac{V(E)}{2\hbar}}\bra\det(E-H)\ket,
\end{aligned} 
\end{equation}
$S_n(\xi)$ can be computed from the connected correlators of the operator $X=\Tr\log(E-H)$
\cite{Eynard:2007kz}
\begin{equation}
\begin{aligned}
 S_{n}(\xi)=\sum_{2g-1+m=n}\frac{1}{m!}\bra X^m\ket_g.
\end{aligned} 
\end{equation}
As demonstrated in \cite{Saad:2019lba}, these correlators can be computed systematically
by the topological recursion.\footnote{
Our normalization of the spectral curve $y=\hf \sin(2z)=z+\mathcal{O}(z^3)$ is
the same as the Airy curve $y=z$ near $z=0$. Thus
the first few orders of resolvent  $\sum_{g}\hbar^{2g-2+n}W_{g,n}(z_1,\cdots,z_n)$
have the same coefficients as the Airy case
\begin{equation}
\begin{aligned}
 W_{0,1}(z)=2zy(z),\quad W_{0,2}(z_1,z_2)=\frac{1}{(z_1-z_2)^2},\quad
W_{0,3}(z_1,z_2,z_3)=\frac{1}{2z_1^2z_2^2z_3^2}.
\end{aligned} 
\end{equation}
For the $g=1$ corrections we find
\begin{equation}
\begin{aligned}
 W_{1,1}(z)=\frac{3+2z^2}{48z^4},\quad
W_{1,2}(z_1,z_2)=\frac{5z_1^4+3z_1^2z_2^2+5z_2^4+4(z_1^2z_2^4+z_1^4z_2^2)+2z_1^4z_2^4}{32z_1^6z_2^6}.
\end{aligned} 
\end{equation}
}
For instance $S_1(\xi)$ comes from the cylinder amplitude
\begin{equation}
\begin{aligned}
 S_1(\xi)=\hf \bra X^2\ket_{g=0}=-\hf\log(2\rt{\xi}).
\end{aligned} 
\end{equation}
Then the order $\mathcal{O}(\hbar^0)$
term of the integral \eqref{eq:BA-lap}
is obtained by evaluating the Gaussian integral around the saddle point
\begin{equation}
\begin{aligned}
 G_1(\la)=S_1(\xi_*)-\hf\log\bigl[-S''_0(\xi_*)\bigr]=-\hf\log\cos(2\rt{\xi_*}).
\end{aligned} 
\end{equation}
One can check that this reproduces the result in \eqref{eq:Gn-result}.

One can easily generalize this calculation to higher order corrections. 
To do this we set
\begin{equation}
\begin{aligned}
 \xi-\xi_*=\rt{\hbar}\phi,
\end{aligned} 
\end{equation}
and perform the integral of $\phi$ perturbatively
by the Wick contraction
with respect to the Gaussian measure around the saddle point $\xi_*$
\begin{equation}
\begin{aligned}
\exp\Biggl(\sum_{n=2}^\infty \hbar^{n-1}G_n(\la)\Biggr)= e^{-\frac{S_0(\xi_*)}{\hbar}-S_1(\xi_*)}
\left\bra\exp\Biggl(\sum_{n=0}^\infty \hbar^{n-1}S_n(\xi_*+\rt{\hbar}\phi)\Biggr)\right\ket
\end{aligned} 
\label{eq:wick}
\end{equation} 
where $\bra \phi^{2m}\ket$ is given by
\begin{equation}
\begin{aligned}
 \bra \phi^{2m}\ket=\frac{\int d\phi e^{\hf S''_0(\xi_*)\phi^2}\phi^{2m}}
{\int d\phi e^{\hf S''_0(\xi_*)\phi^2}}=\frac{(2m-1)!!}{[-S''_0(\xi_*)]^m}.
\end{aligned} 
\end{equation}

Let us compute $G_2(\la)$ using this formalism. At this order we need $S_2(\xi)$, 
which is easily obtained from the topological recursion as
\begin{equation}
\begin{aligned}
 S_2(\xi)=\bra X\ket_{g=1}+\frac{1}{3!}\bra X^3\ket_{g=0}=-\frac{5}{48z^3}-\frac{1}{24z},
\end{aligned} 
\end{equation}
where $z$ is the uniformization coordinate defined in \eqref{eq:E-xi}.
From our general formula \eqref{eq:wick}, $G_2(\la)$ is given by 
\begin{equation}
\begin{aligned}
 G_2(\la)&=S_2(\xi_*)+\hf\left[\frac{S^{(3)}_0(\xi_*)}{3!}\right]^2\bra \phi^6\ket
+\frac{S^{(4)}_0(\xi_*)}{4!}\bra \phi^4\ket+\frac{S^{(3)}_0(\xi_*)S_1'(\xi_*)}{3!}\bra \phi^4\ket\\
&\quad +\hf\left(S_1'(\xi_*)^2+S''_1(\xi_*)\right)\bra \phi^2\ket.
\end{aligned} 
\end{equation}
One can check that this reproduces the result in \eqref{eq:Gn-result}.
We can in principle compute $G_n(\la)$ up to any desired order using this
formalism.

%%%
\subsubsection{WKB expansion of $\rho(E)$}
%%%
We can repeat the same analysis in the previous subsection for
the eigenvalue density $\rho(E)$.
It turns out that the
't Hooft expansion of $Z_\JT$ is related to the WKB expansion
of $\rho(E)$ in the forbidden region $E<0$.

Let us consider the WKB expansion of $\rho(E)$
\begin{equation}
\begin{aligned}
 \rho(E)
=\exp\left[\sum_{n=0}^\infty \hbar^{n-1}\mathcal{S}_n(z)\right].
\end{aligned} 
\end{equation}
$\mathcal{S}_0(z)$ and $\mathcal{S}_1(z)$ are given by \cite{Saad:2019lba}
\begin{equation}
\begin{aligned}
 \mathcal{S}_0(z)&=-V_{\text{eff}}(-z^2),\qquad
\mathcal{S}_1(z)=-\log\left(\frac{8z^2}{\pi}\right).
\end{aligned} 
\end{equation}
As discussed in \cite{Saad:2019lba}, $\mathcal{S}_{n\geq2}(z)$ is written as
some combination of the connected correlator of the operator $Y$
\begin{equation}
\begin{aligned}
 \mathcal{S}_n(z)=\sum_{2g-1+m=n}\frac{1}{m !}\bra Y^m\ket_{g}
\end{aligned} 
\end{equation}
where $Y$ is given by
\begin{equation}
\begin{aligned}
 Y=\Tr\log (E(z)-H)-\Tr \log(E(-z)-H).
\end{aligned} 
\end{equation}
Here the sign of $z$ in $E(\pm z)$ distinguishes the two sheets of the spectral curve.
In other words, $Y$ is defined by integrating the resolvent from $-z$ to $+z$.
Again, one can compute $\mathcal{S}_n(z)$ systematically from the topological recursion.
For instance $\mathcal{S}_2(z)$ is given by
\begin{equation}
\begin{aligned}
\mathcal{S}_2(z)&=\bra Y\ket_{g=1}+\frac{1}{3!}\bra Y^3\ket_{g=0}
=-\frac{17}{24z^3}-\frac{1}{12z}.
\end{aligned} 
\end{equation}
One can check that the saddle point approximation of the integral
\begin{equation}
\begin{aligned}
 Z_\JT&=\int_{-\infty}^\infty dE \rho(E)e^{-\bt E} 
=\int_{-\infty}^\infty d\xi \rho(-\xi)\,e^{\frac{\xi\la}{\hbar}}
\end{aligned} 
\end{equation}
correctly reproduces the free energy $\cF_{n}(\la)$ in the 't Hooft limit in \eqref{eq:CF-n}.
This computation is completely parallel to that
in the previous subsection \ref{sec:wkb-psi}, so we will not repeat it here.

%%%%%%%%%%%%%%%%%%%%%%%%%%%%%%%%%%%%%%%%%%%%%%%%%%%%%%%%%%%%%%%%%%%%%%%%
\section{\mathversion{bold}Numerical analysis of
         $\rho(E)$ and $\psi(E)$}\label{sec:num}
%%%%%%%%%%%%%%%%%%%%%%%%%%%%%%%%%%%%%%%%%%%%%%%%%%%%%%%%%%%%%%%%%%%%%%%%

In this section we will numerically study the behavior of $\rho(E)$
and $\psi(E)$ as a function of $E$.

\begin{figure}[thb]
\centering
\includegraphics[width=8cm]{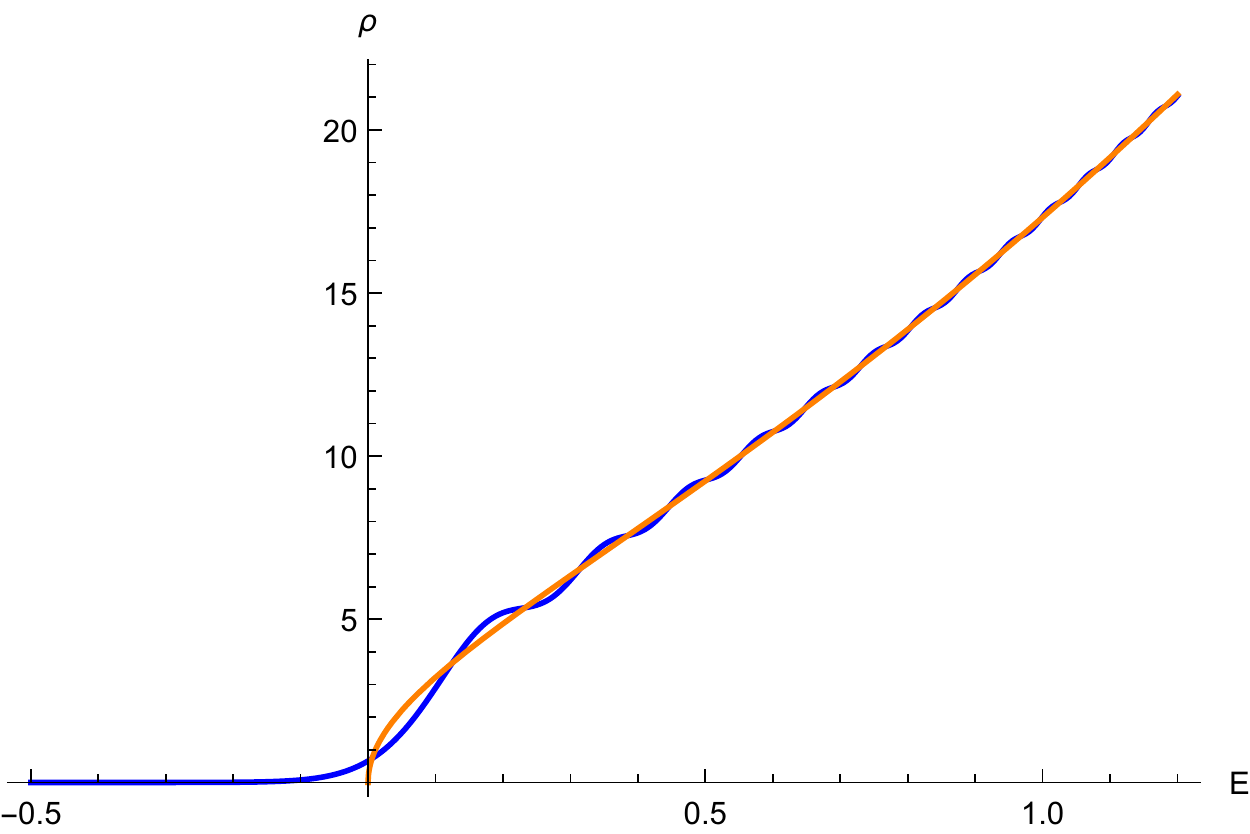}
  \caption{Plot of $\rho(E)$ for $\hbar=1/30$. 
The blue curve represents $\rho(E)$ in the approximation \eqref{eq:rho-approx}
while the orange curve represents the genus-zero 
eigenvalue density $\rho_0(E)$ in \eqref{eq:rho0}.}
  \label{fig:rho}
\end{figure}

Let us consider the integral representation of $\rho(E)$
given by the inverse Laplace transform of $Z_\JT(\bt)$
\begin{equation}
\begin{aligned}
 \rho(E)=\int_C\frac{d\bt}{2\pi\ri}Z_\JT(\bt)e^{\bt E}.
\end{aligned} 
\label{eq:inv-lap}
\end{equation}
Here we take the contour $C$ to be homotopic to the Airy contour. 
We will approximate this integral by keeping the free energy $\cF_n(\la)$ up to $n=2$
obtained in \eqref{eq:F0-period} and \eqref{eq:CF-n}
\begin{equation}
\begin{aligned}
 \rho(E)\approx \int_C\frac{d\la}{2\pi\ri \hbar}\exp\left[\frac{\la E+\cF_0(\la)}{\hbar}
+\cF_1(\la)+\hbar \cF_2(\la)\right].
\end{aligned} 
\label{eq:rho-approx}
\end{equation}
This truncation might be justified when the coupling $\hbar$ is small $\hbar\ll1$.
To avoid the cut of $\cF_n(\la)$ running from $\la=1$ to $\la=+\infty$ 
along the real axis, we choose the contour $C$ to cross
the real axis in the region $0<\text{Re}(\la)<1$.
In practice,
in order to evaluate the integral numerically
we choose $C$ as a union of three straight segments
\begin{equation}
\begin{aligned}
 C=[e^{-\frac{\pi\ri}{3}}\infty,e^{-\frac{\pi\ri}{3}}]\cup
[e^{-\frac{\pi\ri}{3}},e^{\frac{\pi\ri}{3}}]\cup
[e^{\frac{\pi\ri}{3}},e^{\frac{\pi\ri}{3}}\infty]
\end{aligned} 
\label{eq:C-seg}
\end{equation}
and use \texttt{NIntegrate} in \texttt{Mathematica} to evaluate the integral.
In Fig.~\ref{fig:rho} we show the numerical plot of the integral \eqref{eq:rho-approx}.
As expected, $\rho(E)$ approaches the genus-zero value $\rho_0(E)$ in the allowed region $E>0$.
It turns out that the genus-zero part comes from the integral around the origin $\la=0$.
Although our integration contour \eqref{eq:C-seg} does not encircle the origin,
we can deform the contour to pick up the contribution around $\la=0$.
However, we should emphasize that the contour \eqref{eq:C-seg}
is completely fixed in the actual numerical computation of the integral \eqref{eq:rho-approx}.
Near $\la=\hbar\bt=0$, we can go back to the original expression 
\eqref{eq:inv-lap} using $\bt$ as the integration variable.
In the limit $\hbar\to0$ with fixed $\bt$, only $\cF_1(\la)$
and the first term $\cF_2(\la)=\frac{1}{\la}+\mathcal{O}(\la^0)$ in the small $\la$ expansion of $\cF_2(\la)$ survive
\begin{equation}
\begin{aligned}
 \lim_{\hbar\to0}\sum_{n=0}^\infty \hbar^{n-1}\cF_n(\hbar\bt)=
-\log(2\rt{\pi}\hbar\bt^{3/2})+\frac{1}{\bt}+\mathcal{O}(\hbar^0).
\end{aligned} 
\end{equation}
Note that this is the same as the first two terms in the high temperature 
expansion \eqref{eq:high-free}.
It is interesting that the genus-zero term $1/\bt$ in the original
expansion \eqref{eq:high-free} becomes a part of $\cF_2(\la)$ after taking the 't Hooft limit.
Put differently, in order to reproduce the genus-zero part $\rho_0(E)$ numerically
we have to include $\cF_2(\la)$ in the approximation \eqref{eq:rho-approx}.
Then the contribution around $\bt=0$ is evaluated as
\begin{equation}
\begin{aligned}
 \rho(E)\approx \oint_{\bt=0}\frac{d\bt}{2\pi\ri}\frac{e^{\bt E+\frac{1}{\bt}}}{2\rt{\pi}\hbar\bt^{3/2}}=\frac{E^\qu}{2\rt{\pi}\hbar}I_\hf(2\rt{E}).
\end{aligned} 
\label{eq:bt0-int}
\end{equation}
Using the explicit form of the modified Bessel function
\begin{equation}
\begin{aligned}
 I_\hf(z)=\rt{\frac{2}{\pi z}}\sinh(z),
\end{aligned} 
\end{equation}
one can see that \eqref{eq:bt0-int} reproduces the genus-zero eigenvalue density $\rho_0(E)$ in \eqref{eq:rho0}.

Next consider the difference between $\rho(E)$ and $\rho_0(E)$
\begin{equation}
\begin{aligned}
 \rho_{\text{np}}(E)=\rho(E)-\rho_0(E).
\end{aligned} 
\end{equation}
We can estimate this difference by the saddle point approximation of \eqref{eq:rho-approx}.
When $E>0$, we can pick up the contribution of two saddle points on the imaginary axis
of complex $\la$-plane
\begin{equation}
\begin{aligned}
 \la_\pm=\pm i\sinh(2\rt{E}),
\end{aligned} 
\label{eq:la-pm}
\end{equation}
by deforming the contour $C$ within the homotopy class of Airy contour.
Adding the contributions of two saddle points \eqref{eq:la-pm} we find
\begin{equation}
\begin{aligned}
 \rho_{\text{np}}(E)\approx -\frac{1}{4\pi E}\cos\left[\frac{2\rt{E}\cosh(2\rt{E})
-\sinh(2\rt{E})}{2\hbar}\right],\quad(E>0),
\end{aligned} 
\label{eq:rho-np}
\end{equation}
where the prefactor comes from the Gaussian integral around the saddle points.
This agrees with the result of \cite{Saad:2019lba} obtained from a 
different method. 
\begin{figure}[thb]
\centering
\includegraphics[width=8cm]{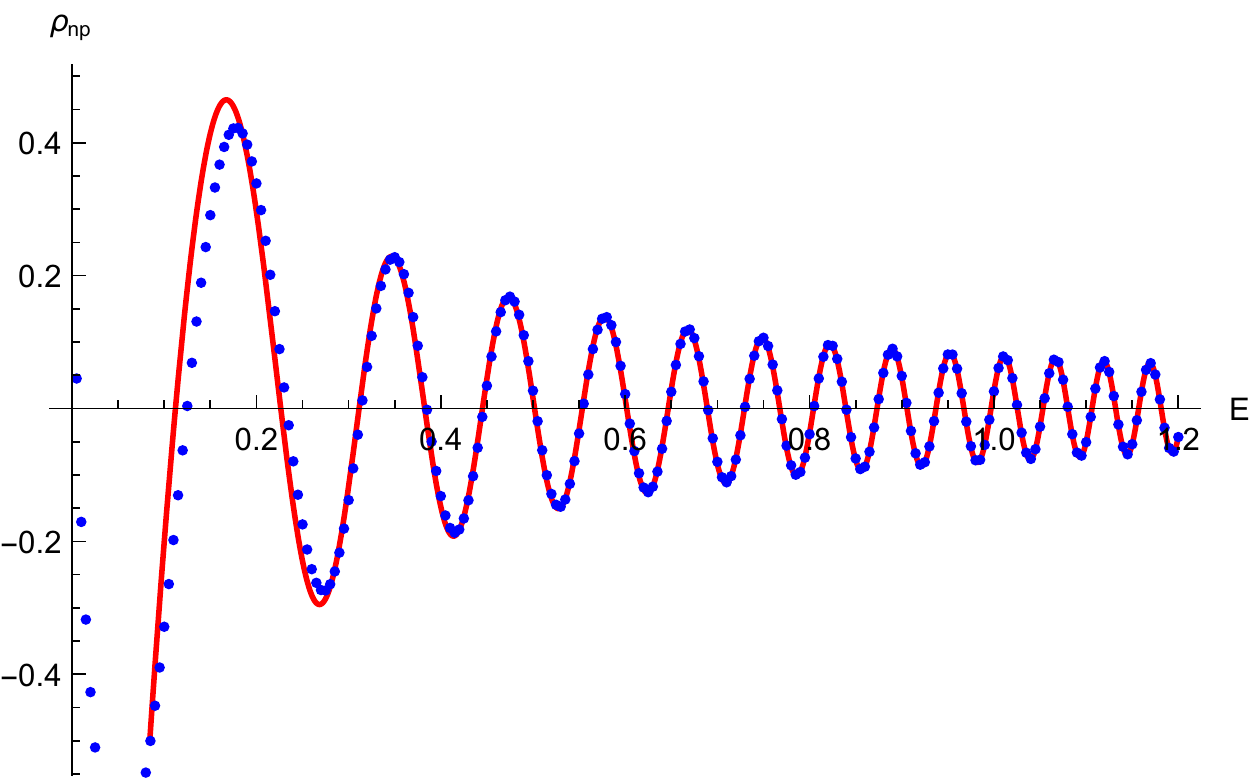}
  \caption{Plot of $\rho_{\text{np}}(E)$ for $\hbar=1/30$.
Blue dots represent the  numerical value of $\rho_{\text{np}}(E)$ obtained from \eqref{eq:rho-approx} while the red curve
is the plot of analytic expression in \eqref{eq:rho-np}.}
  \label{fig:rho-np}
\end{figure}
In Fig.~\ref{fig:rho-np} we show the plot of $\rho_{\text{np}}(E)$.
One can see that the numerical value of $\rho_{\text{np}}(E)$ 
fits nicely with the analytic expression in \eqref{eq:rho-np}.

\begin{figure}[thb]
\centering
\includegraphics[width=8cm]{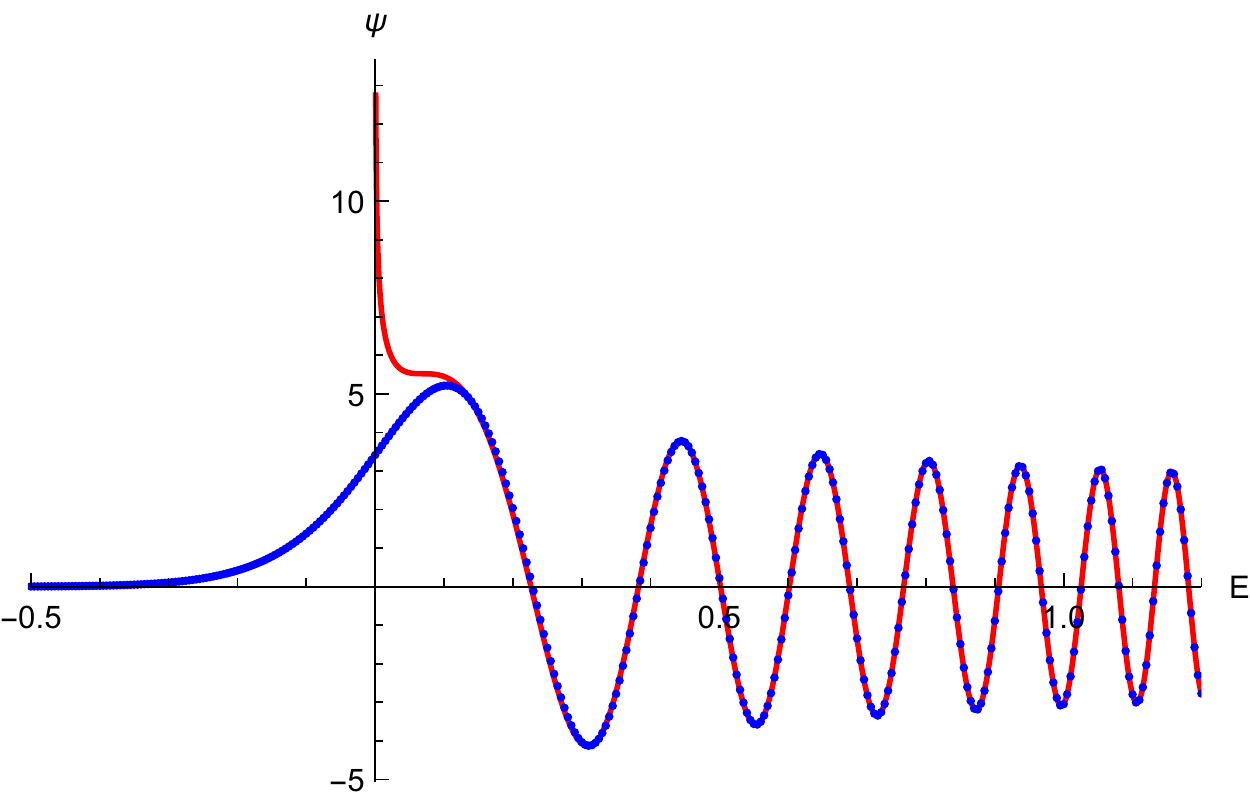}
  \caption{Plot of $\psi(E)$ for $\hbar=1/30$.
Blue dots represent the  numerical value of $\psi(E)$ obtained from \eqref{eq:psi-approx} 
while the red curve
is the plot of analytic expression in \eqref{eq:psi-osc}.}
  \label{fig:psi}
\end{figure}

In a similar manner we can numerically compute the BA function $\psi(E)$
in the approximation of keeping $G_0(\la)$ and $G_1(\la)$ in
\eqref{eq:Gn-result} in the 't Hooft expansion
of $\h{\psi}(\la)$ \eqref{eq:hat-psi}
\begin{equation}
\begin{aligned}
 \psi(E)\approx \int_C\frac{d\la}{4\pi\hbar\ri}(1-\la^2)^{-\qu}\exp\left(\frac{\la E+\cF_0(\la)}{2\hbar}\right).
\end{aligned} 
\label{eq:psi-approx}
\end{equation} 
In this case we do not have to include $G_2(\la)$ for the purpose of numerical analysis
since $G_2(\la)=\mathcal{O}(\la)$ in the small $\la$ expansion and hence
there is no non-trivial contribution from $\la=0$.
Again, in the allowed region $E>0$ there are two saddle points $\la_\pm$ in \eqref{eq:la-pm}.
Adding the contributions of these saddle points we find
\begin{equation}
\begin{aligned}
 \psi(E)\approx \frac{1}{\rt{\pi\hbar}E^\qu}\cos\left[
\frac{2\rt{E}\cosh(2\rt{E})
-\sinh(2\rt{E})}{4\hbar}-\frac{\pi}{4}\right],\quad(E>0).
\end{aligned} 
\label{eq:psi-osc}
\end{equation}
In Fig.~\ref{fig:psi} we show the plot of $\psi(E)$.
One can see that the numerical value of $\psi(E)$ agrees well with
the saddle point result \eqref{eq:psi-osc} in the allowed region $E>0$.

Let us consider the behavior of $\psi(E)$ in the forbidden region $E<0$.
Naively, when $E<0$ there is a saddle point
\begin{equation}
\begin{aligned}
 \la_*=\sin(2\rt{-E})
\end{aligned} 
\label{eq:la-*}
\end{equation}
and it contributes to $\psi(E)$ as
\begin{equation}
\begin{aligned}
 \psi(E)\approx \exp\left[-\frac{V_{\text{eff}}(E)}{2\hbar}\right],\quad(E<0),
\end{aligned} 
\end{equation}
where the effective potential $V_{\text{eff}}(E)$ is given by \eqref{eq:V-eff}.
It is argued in \cite{Saad:2019lba}
that this model is non-perturbatively unstable since
$V_{\text{eff}}(E)$ is not positive definite and $\psi(E)$ blows up as $E\to-\infty$.

\begin{figure}[thb]
\centering
\includegraphics[width=8cm]{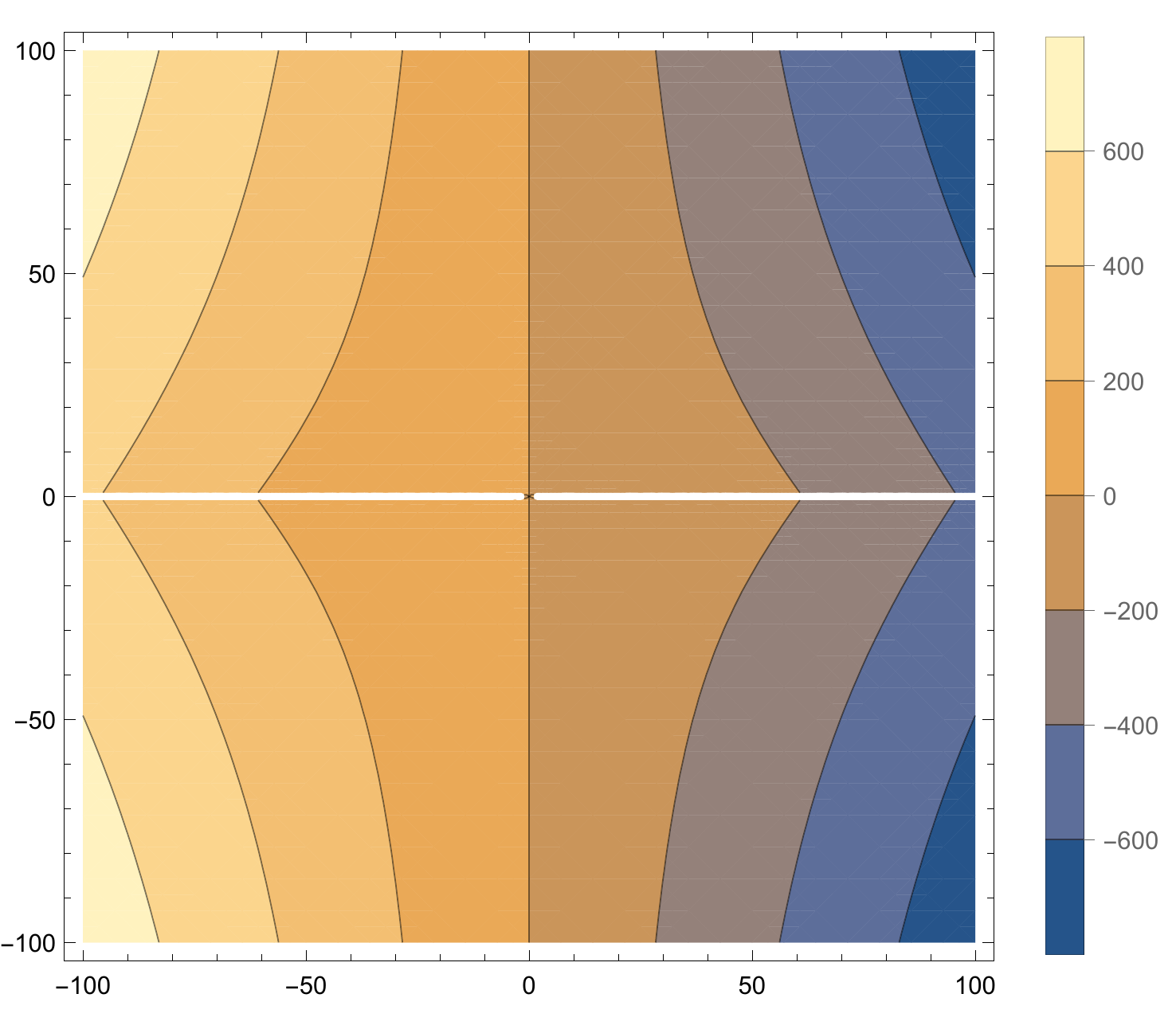}
  \caption{Plot of the real part of $\cF_0(\la)$ on the complex $\la$-plane.
}
  \label{fig:Re-f0}
\end{figure}

However, we do not see this pathological behavior in the numerical plot 
of $\psi(E)$ in Fig.~\ref{fig:psi}.
As we can see from Fig.~\ref{fig:Re-f0}, $\text{Re}[\cF_0(\la)]$ is negative
in the region $\text{Re}(\la)>0$. Thus, the real part of the
leading term $E\la+\cF_0(\la)$
in the WKB expansion \eqref{eq:psi-approx} 
is negative for $E<0$ with an appropriate choice of contour
$C$.  This suggests that the integral representation of $\psi(E)$ in \eqref{eq:psi-Lap}
and its approximation \eqref{eq:psi-approx} 
are convergent and well-defined in the region $E<0$ as long as the contour
$C$ lies on the right half $\text{Re}(\la)>0$ of the complex $\la$-plane,
under the condition that $C$ crosses the real axis in the region $0<\la<1$
to avoid the cut of $\cF_0(\la)$.

Now suppose that we decrease the value of $E$ from $E=0$ toward the negative $E$ direction.
At the begging $E\sim0$ 
the saddle point $\la_*$ in \eqref{eq:la-*} lies on the positive half plane $\la_*>0$,
but $\la_*$ turns negative at $E=-\frac{\pi^2}{4}$ and it ceases to contribute to the
integral below $E=-\frac{\pi^2}{4}$. This suggests that $V_{\text{eff}}(E)$ in \eqref{eq:V-eff}
cannot be trusted for $E<-\frac{\pi^2}{4}$.
It is tempting to speculate that this model is actually non-perturbatively \textit{stable}.
It would be very interesting to understand the non-perturbative
instability discussed in \cite{Saad:2019lba} better.

%%%%%%%%%%%%%%%%%%%%%%%%%%%%%%%%%%%%%%%%%%%%%%%%%%%%%%%%%%%%%%%%%%%%%%%%
\section{Comment on the spectral form factor}\label{sec:SFF}
%%%%%%%%%%%%%%%%%%%%%%%%%%%%%%%%%%%%%%%%%%%%%%%%%%%%%%%%%%%%%%%%%%%%%%%%

One can easily generalize our expression of the macroscopic loop operator 
$Z_\JT=\Tr(e^{\bt Q}\Pi)$ to the case of arbitrary numbers of boundaries
by applying the general formula in \cite{Banks:1989df}
to the JT gravity case $t_n=\ga_n$.
Of particular interest is the connected correlator of two macroscopic loops
and its analytic continuation known as the spectral form factor.
The spectral form factor is extensively studied in the literature
as a useful diagnostics of the quantum chaos of the SYK model
and its bulk gravity dual \cite{Garcia-Garcia:2016mno,Cotler:2016fpe,Saad:2018bqo,saad}.

The connected two-loop correlator is written as \cite{Banks:1989df}
\begin{equation}
\begin{aligned}
 \bra Z(\bt_1)Z(\bt_2)\ket_{\text{conn}}&=\Tr \big(e^{\bt_1Q}(1-\Pi)e^{\bt_2Q}\Pi\big),
\end{aligned} 
\label{eq:two-loop}
\end{equation}
and the spectral form factor is obtained by an analytic continuation
of the correlator
\begin{equation}
\begin{aligned}
g(\bt,t)= \bra Z(\bt+\ri t)Z(\bt-\ri t)\ket_{\text{conn}}.
\end{aligned} 
\end{equation}
As a function of $t$, $g(\bt,t)$
exhibits characteristic features called ramp and plateau.
These features naturally correspond to the following decomposition of \eqref{eq:two-loop}
\begin{equation}
\begin{aligned}
 g(\bt,t)&=\Tr \big(e^{2\bt Q}\Pi\big)
-\Tr\Big( e^{(\bt+\ri t)Q}\Pi e^{(\bt-\ri t)Q}\Pi\Big)\\
&=\bra Z(2\bt)\ket-\Tr \Big(e^{(\bt+\ri t)Q}\Pi e^{(\bt-\ri t)Q}\Pi\Big).
\end{aligned} 
\label{eq:jt-ssf}
\end{equation}
The first term is independent of $t$ and it sets the value of plateau.
On the other hand, the second term is a non-trivial function
of $t$ and it is expected that this term gives rise to the linearly growing ramp.
It is interesting to show this explicitly for JT gravity.
(See appendix \ref{app:airy} for the computation of the spectral form factor 
in the Airy case).
It would also be interesting to consider the bulk gravity picture of plateau.
The first term of \eqref{eq:jt-ssf} 
might be interpreted on the bulk gravity side as 
a geometry where the two boundary circles are merged into a single boundary.
Such a geometry was considered before in the context of 2d gravity
(see Fig.~20 in \cite{Ginsparg:1993is}), but its status in the bulk geometry
is not clear as mentioned in \cite{Ginsparg:1993is}.

%%%%%%%%%%%%%%%%%%%%%%%%%%%%%%%%%%%%%%%%%%%%%%%%%%%%%%%%%%%%%%%%%%%%%%%%
\section{Conclusions and outlook \label{sec:discussion}}
%%%%%%%%%%%%%%%%%%%%%%%%%%%%%%%%%%%%%%%%%%%%%%%%%%%%%%%%%%%%%%%%%%%%%%%%

In this paper we have seen that the
partition function of JT gravity $Z_\JT(\bt)=\bra Z(\bt)\ket$
is written as the expectation value of the macroscopic loop operator
$\Tr (e^{\bt Q}\Pi)$ in the matrix model
of 2d gravity in the closed sting background $t_n=\ga_n$ \eqref{eq:gammavalue1}.
By deforming this background by the two parameters $(t_0,t_1)$,
one can utilize the KdV equation to compute the genus expansion of $Z_\JT$ in a very
efficient way.
We have also shown that the low temperature expansion of $Z_\JT$
as well as its 't Hooft limit \eqref{eq:tHooft}
can be obtained systematically.
By evaluating the inverse Laplace transformation numerically,
we have confirmed the oscillating behavior of $\rho(E)$ and $\psi(E)$
in the region $E>0$ as discussed in \cite{Saad:2019lba}. Interestingly, 
the oscillating cosine term arises by adding the contributions of two saddle points
\eqref{eq:la-pm}. 
On the other hand, we do not see any evidence of the pathological behavior
of $\rho(E)$ and $\psi(E)$
in the region $E<0$ within our approximation.
It would be very interesting to understand the non-perturbative instability discussed in 
\cite{Saad:2019lba}. 
It is desirable to perform more detailed numerical analysis
of $\psi(E)$ along the lines of \cite{Maldacena:2004sn}.\footnote{
We would like to thank Douglas Stanford for emphasizing this point.
}

There are many open questions and interesting future directions.
First, it is interesting to understand the physical meaning of the background $t_n=\ga_n$
corresponding to
JT gravity. Naively one can imagine that the asymptotic $AdS_2$ 
is ``built'' by this background. To see this more quantitatively, it would be useful to
study the Kontsevich's matrix Airy integral \cite{Kontsevich} corresponding to the background
$t_n=\ga_n$. In the modern interpretation
\cite{Maldacena:2004sn,Gaiotto:2003yb,Bertola},
the Kontsevich's model and
Witten's 
topological gravity \cite{Witten} are related by the open/closed duality;
the Kontsevich's model arises as the open string theory on the FZZT branes
while the closed string background $t_n$ is obtained by 
replacing the insertion of FZZT branes
with the deformation of matrix model potential.
It would be interesting to understand the configuration of background FZZT branes
corresponding to $t_n=\ga_n$ (see also footnote \ref{footnote}).

It is very important to understand the analytic properties of the genus expansion 
of $Z_\JT$ and its non-perturbative completion.
Apparently, the 't Hooft expansion of the free energy becomes singular at $\la=1$, and 
the analytic form of $\cF_n(\la)$ in \eqref{eq:CF-n} can be trusted only in the region $\la<1$
when $\la$ is real.
It is interesting to understand what happens at $\la=1$ (or $\bt=\hbar^{-1}$).
Also, it is very important to
see if JT gravity is non-perturbatively well-defined.
One possible avenue is to study the string equation for $u(x)$
in the background $t_n=\ga_n$, which we will discuss briefly
in appendix \ref{app:string}. 

In section \ref{sec:low-rho} we have constructed
the full eigenvalue density $\rho(E)$ as a low energy expansion
in the limit \eqref{eq:small-E-lim} starting from the Airy case $\rho_{\text{Airy}}(E)$.
It would be very significant if we can find the exact eigenvalue density $\rho(E)$. 
It is argued in \cite{Garcia-Garcia:2017pzl,Berkooz:2018jqr} that the
eigenvalue density of the SYK model is closely related to the $q$-Hermite polynomials.
It would be interesting to see if the double scaling limit of the $q$-Hermite polynomials
has some connection to the exact eigenvalue density $\rho(E)$ of the JT gravity case.

In section \ref{sec:SFF} we have briefly commented on the spectral form factor.
Using the result of \cite{Banks:1989df} it is straightforward to write down the
connected correlator of two macroscopic loops \eqref{eq:two-loop}.
It would be interesting to compute it at least in the genus expansion.
To this end, we need to know not only the diagonal matrix element
$W=\bra x|e^{\bt Q}|x\ket$ but also the non-diagonal part $\bra x|e^{\bt Q}|y\ket$.
Fortunately, it is known \cite{Iliev} that the non-diagonal matrix element 
$\bra x|e^{\bt Q}|y\ket$ is written as some combination of the
derivatives of tau-function, hence
it is possible to generalize
the method of KdV equation in our paper
to the computation of multi-boundary correlators.
We will report on the computation of multi-boundary correlators
elsewhere \cite{OkuyamaSakai}.
The result of the spectral form factor in the Airy case
\eqref{eq:t-del-g} indicates that in the double-scaled matrix model
the time scale of the transition to plateau 
diverges as $\bt\to0$, 
which deserves further investigation. 

Finally, it is interesting to extend our approach to 
more general settings, including JT supergravity \cite{Stanford:2019vob},
adding gauge fields to the bulk theory \cite{Iliesiu:2019lfc},
and a possible analytic continuation to the 2d de Sitter space 
\cite{Maldacena:2019cbz,Cotler:2019nbi}, to name a few.

\vskip8mm
\acknowledgments
We would like to thank Peter Zograf for sharing his data of
$V_{g,1}(b)$ up to $g=20$.
KO would like to thank Hiroyuki Fuji for useful discussion
during the Chubu Summer School 2019, and Douglas Stanford for correspondence.
This work was supported in part by JSPS KAKENHI Grant
Nos.~26400257, 16K05316, 19K03845 and 19K03856,
and JSPS Japan-Russia Research Cooperative Program.

%%%%%%%%%%%%%%%%%%%%%%%%%%%%%%%%%%%%%%%%%%%%%%%%%%%%%%%%%%%%%%%%%%%%%%%%
\appendix
%%%%%%%%%%%%%%%%%%%%%%%%%%%%%%%%%%%%%%%%%%%%%%%%%%%%%%%%%%%%%%%%%%%%%%%%
\section{Airy case}\label{app:airy}
%%%%%%%%%%%%%%%%%%%%%%%%%%%%%%%%%%%%%%%%%%%%%%%%%%%%%%%%%%%%%%%%%%%%%%%%
In this appendix we summarize the result in the Airy case,
where the spectral curve is given by
\begin{equation}
\begin{aligned}
 y=\rt{\xi},
\end{aligned} 
\end{equation}
and the corresponding classical eigenvalue density is
\begin{equation}
\begin{aligned}
 \rho_0(E)=\frac{\rt{E}}{\pi\hbar}.
\end{aligned} 
\label{eq:rho0-airy}
\end{equation}
This is realized by a double scaling limit of the Gaussian
matrix model by
zooming in on the edge of the Wigner semi-circle
(see
\cite{Maldacena:2004sn} and references therein).
In this case
$u(x)=x$ and $Q$ is given by
\begin{equation}
\begin{aligned}
 Q=\hbar^2\del_x^2+x.
\end{aligned} 
\end{equation}
In this appendix we will use the normalization $t_0=x$, which differs from
\eqref{eq:rescaledvar} by a factor of $\hbar$.
The BA function obeying $(Q+E)\psi (E)=0$ is given by
the Airy function
\begin{equation}
\begin{aligned}
\psi(E)= \bra x|E\ket=\hbar^{-\frac{2}{3}}\text{Ai}\bigl[-\hbar^{-\frac{2}{3}}(E+x)\bigr].
\end{aligned} 
\label{eq:BA-airy}
\end{equation}
One can show that $\psi(E)$ in \eqref{eq:BA-airy} is normalized as
\begin{equation}
\begin{aligned}
 \bra E|E'\ket=\int_{-\infty}^\infty dx \bra E|x\ket\bra x|E'\ket=\cob(E-E').
\end{aligned} 
\end{equation}

Now let us consider the one-point function of macroscopic loop operator
\begin{equation}
\begin{aligned}
 \bra Z(\bt)\ket&=
\int_{-\infty}^0 dx \bra x|e^{\bt Q}|x\ket=
\int_{-\infty}^\infty dE e^{-\bt E} \rho_{\text{Airy}}(E),
\end{aligned} 
\label{eq:airy-one}
\end{equation}
where the eigenvalue density $\rho_{\text{Airy}}(E)$ is given by
\begin{equation}
\begin{aligned}
\rho_{\text{Airy}}(E)=\int_{-\infty}^0 dx \bra x|E\ket^2.
\end{aligned} 
\label{eq:rho-airy-int}
\end{equation}
Using the expression of BA function $\bra x|E\ket$ in \eqref{eq:BA-airy},
one can show that \eqref{eq:rho-airy-int} reproduces 
the eigenvalue density in \eqref{eq:rho-airy}.
This defines a non-perturbative completion of the classical
eigenvalue density \eqref{eq:rho0-airy}.
We can evaluate the integral in \eqref{eq:airy-one} and find
\begin{equation}
\begin{aligned}
 \bra Z(\bt)\ket=\frac{e^{\frac{\hbar^2\bt^3}{12}}}{2\rt{\pi}\hbar\bt^{3/2}}.
\end{aligned} 
\label{eq:Z-airy}
\end{equation}
This can be thought of as the generating function for
the intersection numbers $\bra\psi_1^{3g-2}\ket_{g,1}$.

Next consider the connected correlator of two macroscopic loops
\begin{equation}
\begin{aligned}
 \bra Z(\bt_1)Z(\bt_2)\ket_{\text{conn}}&=\Tr \big(e^{\bt_1Q}(1-\Pi)e^{\bt_2Q}\Pi\big)
\end{aligned} 
\end{equation}
where $\Pi$ is the projector
\begin{equation}
\begin{aligned}
 \Pi=\int_{-\infty}^0 dx|x\ket\bra x|.
\end{aligned} 
\end{equation}
The general
$n$-loop amplitude $\bra \prod_{i=1}^n Z(\bt_i)\ket_{\text{conn}}$ has been computed in \cite{Okounkov}
and 
the result for the two-loop correlator reads
\begin{equation}
\begin{aligned}
 \bra Z(\bt_1)Z(\bt_2)\ket_{\text{conn}}&=\frac{e^{\frac{\hbar^2}{12}(\bt_1+\bt_2)^3}}{2\rt{\pi}\hbar(\bt_1+\bt_2)^{3/2}}\text{Erf}\left(
\frac{\hbar}{2}\rt{\bt_1\bt_2(\bt_1+\bt_2)}\right),
\end{aligned} 
\label{eq:airy-conn}
\end{equation}
where $\text{Erf}(z)$ denotes the error function
\begin{equation}
\begin{aligned}
 \text{Erf}(z)=\frac{2}{\rt{\pi}}\int_0^z dte^{-t^2}.
\end{aligned} 
\end{equation}
It is interesting to compare the connected part and the disconnected part
of the two-loop correlator as a function of $\bt$
\begin{equation}
\begin{aligned}
 \bra Z(\bt)^2\ket_{\text{dis}}&=\bra Z(\bt)\ket^2
=\frac{e^{\frac{\hbar^2\bt^3}{6}}}{4\pi\hbar^2\bt^3},\\
\bra Z(\bt)^2\ket_{\text{conn}}&= \frac{e^{\frac{\hbar^2(2\bt)^3}{12}}}{2\rt{\pi}\hbar(2\bt)^{3/2}}
\text{Erf}\Bigl(\frac{\hbar}{\rt{2}}\bt^{3/2}\Bigr).
\end{aligned} 
\label{eq:dis-conn-airy}
\end{equation}
Here we have set $\bt_1=\bt_2=\bt$ for simplicity.
\begin{figure}[thb]
\centering
\includegraphics[width=8cm]{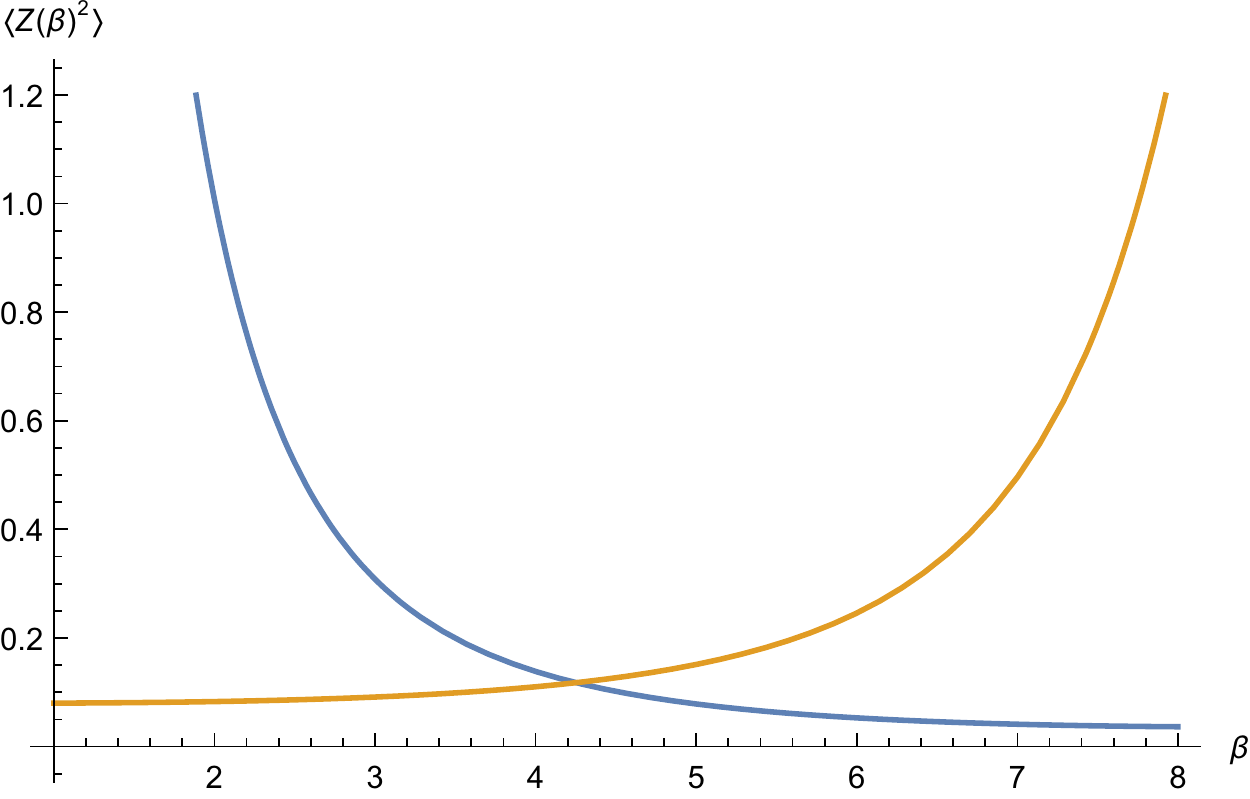}
  \caption{Plot of two-loop correlator in the Airy case for $\hbar=1/10$. 
The blue curve represents the disconnected part $\bra Z(\bt)^2\ket_{\text{dis}}$
while the orange curve represents the connected part $\bra Z(\bt)^2\ket_{\text{conn}}$.
}
  \label{fig:airy-twoloop}
\end{figure}
In Fig.~\ref{fig:airy-twoloop} we show the plot of the two-loop correlator
in \eqref{eq:dis-conn-airy} for $\hbar=1/10$.
One can see that at high temperature (small $\bt$)
the disconnected part (blue curve) is dominant, while at low temperature
(large $\bt$) the connected part (orange curve) becomes dominant.
As we lower the temperature there occurs the exchange of
dominance between the disconnected and the connected part at some critical value 
$\bt=\bt_{\text{crit}}$. 
A similar phenomenon was observed in the coupled SYK model \cite{Maldacena:2018lmt}
and it was interpreted as the Hawking-Page transition on the bulk gravity side.
Since 
both $\bra Z(\bt)^2\ket_{\text{dis}}$ and $\bra Z(\bt)^2\ket_{\text{conn}}$
depend on $\bt$ only through the combination $\hbar^2\bt^3$, 
the critical temperature scales as\footnote{A similar scaling behavior of $\bt_{\text{crit}}$
also appeared in the Gaussian matrix model
before taking the double-scaling limit \cite{Okuyama:2019xvg}.}
\begin{equation}
\begin{aligned}
 \bt_{\text{crit}}\sim \hbar^{-\frac{2}{3}}.
\end{aligned} 
\end{equation}

We can also study the spectral form factor in the 
Airy case by analytically continuing the result in \eqref{eq:airy-conn}
\begin{equation}
\begin{aligned}
 g(\bt,t)=\bra Z(\bt+\ri t)Z(\bt-\ri t)\ket_{\text{conn}}.
\end{aligned} 
\end{equation}
It turns out that the time derivative of $g(\bt,t)$ has a simple form
\begin{equation}
\begin{aligned}
 \del_t g(\bt,t)=\frac{t}{4\pi \bt\rt{\bt^2+t^2}}e^{\frac{1}{6}\hbar^2\bt^3-\hf \bt\hbar^2t^2}.
\end{aligned} 
\label{eq:t-del-g}
\end{equation}
This $\del_t g(\bt,t)$ decays exponentially at large $t$
and the spectral form factor approaches a constant plateau at late times
\begin{equation}
\begin{aligned}
 \lim_{t\to\infty}g(\bt,t)=\frac{e^{\frac{\hbar^2}{12}(2\bt)^3}}{2\rt{\pi}\hbar (2\bt)^{3/2}}
=\bra Z(2\bt)\ket.
\end{aligned} 
\end{equation}
From \eqref{eq:t-del-g} one can read off
the time scale for the transition from ramp to plateau 
as
\begin{equation}
\begin{aligned}
 t_{\text{plateau}}\sim \frac{1}{\hbar\rt{\bt}}.
\end{aligned} 
\label{eq:t-plateau}
\end{equation}
We notice that $t_{\text{plateau}}$ depends on $\bt$.
This is in contrast to the situation 
in the Gaussian matrix model before taking the double scaling limit
where $t_{\text{plateau}}$ is independent of $\bt$ \cite{Okuyama:2018yep}.

It is interesting to observe that the $\bt\to0$ limit of
$g(\bt,t)$ is singular due to the one-loop factor $(\bt_1+\bt_2)^{-3/2}$
in \eqref{eq:airy-conn}. In \cite{Maldacena:2019cbz,Cotler:2019nbi}
the analytically continued two-loop correlator
$\bra\Tr e^{-\ri \ell H}\Tr e^{\ri \ell H}\ket$ in the JT gravity case 
was interpreted as the inner product of Wheeler de Witt wave functions of the 2d de Sitter space.
This inner product naively corresponds to $g(0,\ell)$, which is divergent.
Interpretation of the singularity of $g(\bt,t)$ at $\bt=0$ is unclear at present.

%%%%%%%%%%%%%%%%%%%%%%%%%%%%%%%%%%%%%%%%%%%%%%%%%%%%%%%%%%%%%%%%%%%%%%%%
\section{Partial resummation of the eigenvalue density}\label{app:partial}
%%%%%%%%%%%%%%%%%%%%%%%%%%%%%%%%%%%%%%%%%%%%%%%%%%%%%%%%%%%%%%%%%%%%%%%%

In this appendix we consider a partial resummation of the
low energy expansion of $\rho(E)$.
We observe from \eqref{eq:zh} that
\begin{equation}
\begin{aligned}
 \lim_{h\to0}\tz_{\ell}(h)=1.
\end{aligned} 
\end{equation}
Then it is 
interesting to see what happens if we replace $\tz_\ell(h)\to1$ 
and perform the summation over $\ell$ in the low temperature expansion of $Z_\JT$
in \eqref{eq:Z-tz}.
This replacement leads to
\begin{equation}
\begin{aligned}
 Z_{\text{partial}}(\bt)&=
\frac{e^{\frac{h^2}{12}}}{2\rt{\pi}h}\sum_{\ell=0}^\infty\frac{T^\ell}{\ell!}\tz_\ell(h=0)
=\frac{e^{\frac{\hbar^2\bt^3}{12}+\frac{1}{\bt}}}{2\rt{\pi}\hbar\bt^{3/2}}.
\end{aligned} 
\label{eq:z-partial}
\end{equation}
By the inverse Laplace transformation we find
a simple closed form expression of the eigenvalue density $\rho_{\text{partial}}(E)$
for $Z_{\text{partial}}(\bt)$
\begin{equation}
\begin{aligned}
 \rho_{\text{partial}}(E)&=\int_C\frac{d\bt}{2\pi\ri}e^{E\bt}Z_{\text{partial}}(\bt)\\
&=\frac{1}{\hbar}\text{Im}
\Bigl[\text{Ai}\Big(\hbar^{-\frac{2}{3}}(-E+\ri\hbar)\Big)
\text{Ai}'\Big(\hbar^{-\frac{2}{3}}(-E-\ri\hbar)\Big)\Bigr]\\
&=\frac{1}{\hbar}\text{Im}
\Bigl[\text{Ai}(\eta+\ri \hbar^{\frac{1}{3}})
\text{Ai}'(\eta-\ri \hbar^{\frac{1}{3}})\Bigr],
\end{aligned} 
\label{eq:rho-part}
\end{equation}
which can be thought of as a partial resummation of the expansion
\eqref{eq:rho-eta}.
It turns out that in the classical limit \eqref{eq:rho-part} reduces to
the genus-zero eigenvalue density $\rho_0(E)$ 
of the \textit{full} partition function $Z_\JT$
\begin{equation}
\begin{aligned}
 \lim_{\hbar\to0}\rho_{\text{partial}}(E)=\frac{\sinh(2\rt{E})}{2\pi\hbar}.
\end{aligned} 
\label{eq:rho-class}
\end{equation}
This is expected since $Z_{\text{partial}}(\bt)$ in \eqref{eq:z-partial}
reduces in the limit $\hbar\to0$
to the genus-zero part of $Z_\JT$ in \eqref{eq:ZJT-g}.

\begin{figure}[thb]
\centering
\includegraphics[width=8cm]{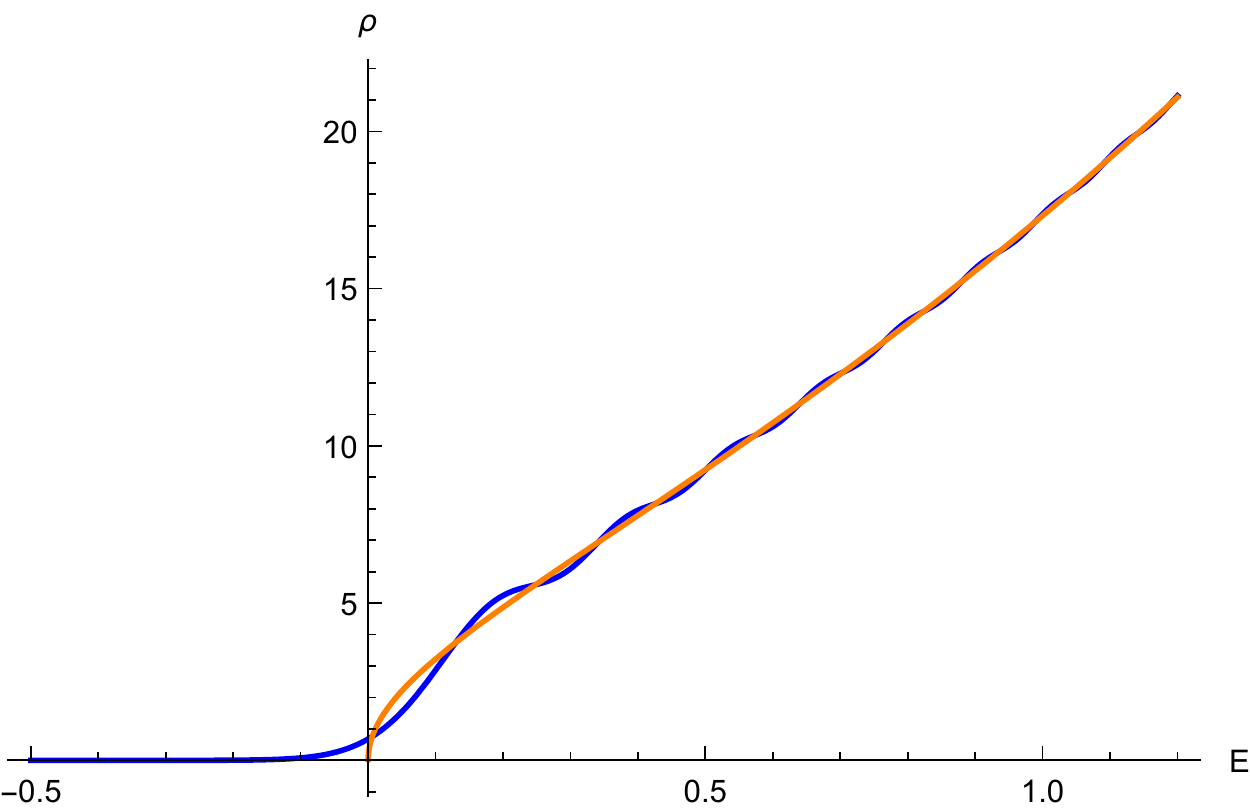}
  \caption{Plot of $\rho_{\text{partial}}(E)$ for $\hbar=1/30$. 
The blue curve represents  $\rho_{\text{partial}}(E)$ in \eqref{eq:rho-part}
while the orange curve represents the genus-zero 
eigenvalue density $\rho_0(E)$ in \eqref{eq:rho0}.}
  \label{fig:rho-part}
\end{figure}
To see that this is indeed the case, 
in Fig.~\ref{fig:rho-part} we show the plot of $\rho_{\text{partial}}(E)$ in \eqref{eq:rho-part} for $\hbar=1/30$. One can see that
$\rho_{\text{partial}}(E)$ agrees with $\rho_0(E)$ in \eqref{eq:rho0}
in the allowed region $E>0$ up to an oscillatory correction.
This implies that the genus-zero part $\rho_0(E)$ \eqref{eq:rho0}
is completely accounted for by $\rho_{\text{partial}}(E)$
and the difference from the true density $\rho(E)$
has only oscillatory contribution in the region $E>0$
\begin{equation}
\begin{aligned}
 \rho(E)-\rho_{\text{partial}}(E)=(\text{oscillatory}),\qquad(E>0).
\end{aligned} 
\end{equation}

%%%%%%%%%%%%%%%%%%%%%%%%%%%%%%%%%%%%%%%%%%%%%%%%%%%%%%%%%%%%%%%%%%%%%%%%
\section{String equation for JT gravity}\label{app:string}
%%%%%%%%%%%%%%%%%%%%%%%%%%%%%%%%%%%%%%%%%%%%%%%%%%%%%%%%%%%%%%%%%%%%%%%%

In this appendix we consider the so-called string equation 
for $u(x)$ (see \cite{DiFrancesco:1993cyw}
for a review).
It is known that the genus-zero relation \eqref{eq:g0-string-eq}
can be promoted to the all-genus string equation \cite{Douglas:1989dd}
\begin{equation}
\begin{aligned}
 {[}P,Q{]}&=\hbar,
\end{aligned} 
\label{eq:string-eq}
\end{equation} 
which arises from the compatibility condition for the following
set of equations obeyed by the BA function
\begin{equation}
\begin{aligned}
 Q\psi=\xi\psi,\quad
P\psi=\hbar\del_\xi\psi.
\end{aligned} 
\end{equation}
Here $Q$ is defined in \eqref{eq:Q}. To find
$P$, we start with the relation
\begin{equation}
\begin{aligned}
 \del_\xi\psi=-\sum_{k=1}^\infty \til{t}_{k}\del_{k-1}\psi,
\end{aligned} 
\label{eq:del-xi-psi}
\end{equation}
where $\til{t}_k$ is defined by
\begin{equation}
\begin{aligned}
 \til{t}_k=t_k-\cob_{k,1}.
\end{aligned}
\label{eq:til-tk} 
\end{equation}
$\del_k\psi$ is given by the KdV flow equation
in the $k$-th direction 
\begin{equation}
\begin{aligned}
 \hbar\del_k\psi=M_k\psi
\end{aligned} 
\label{eq:kdv-flow}
\end{equation}
with $M_k$ being
\begin{equation}
\begin{aligned}
 M_k=\frac{2^{k}}{(2k+1)!!}Q_+^{k+\hf}.
\end{aligned} 
\end{equation}
Here the subscript $+$ of $Q_+^{k+\hf}$ indicates that we truncate the pseudo-differential
operator $Q^{k+\hf}$ to its differential part. In this notation $M$ in 
\eqref{eq:linprob} is written as
$M=M_1=\frac{2}{3}Q_+^{3/2}$.
From \eqref{eq:del-xi-psi} and \eqref{eq:kdv-flow}
we find that $P$ is given by
\begin{equation}
\begin{aligned}
P&=-\sum_{k=1}^\infty \til{t}_kM_{k-1}=
-\sum_{k=1}^\infty \til{t}_k \frac{2^{k-1}}{(2k-1)!!}Q_+^{k-\hf}.
\end{aligned} 
\label{eq:P-diff}
\end{equation}
The compatibility of the flow equation and $Q\psi=\xi\psi$
leads to the following relation
\begin{equation}
\begin{aligned}
\hbar\del_ku=\hbar\del_0\cR_{k+1}= \bigl[M_{k},Q\bigr],
\end{aligned} 
\end{equation}
where we used $u=\del_0^2 F$ and \eqref{eq:FRrel}.
Then the string equation \eqref{eq:string-eq} becomes
\begin{equation}
\begin{aligned}
 \hbar=[P,Q]=-\sum_{k=1}^\infty\til{t}_k[M_{k-1},Q]=-\hbar\sum_{k=1}^\infty\til{t}_k \del_0\cR_k,
\end{aligned} 
\end{equation}
which can be integrated as 
\begin{equation}
\begin{aligned}
 t_0=-\sum_{k=1}^\infty\til{t}_k \cR_k.
\end{aligned} 
\label{eq:qu-string}
\end{equation}
Using $\cR_0=1$ this is more compactly written as
\begin{equation}
\begin{aligned}
 \sum_{k=0}^\infty \til{t}_k \cR_k=0.
\end{aligned} 
\end{equation} 
From the behavior of $\cR_k$ in \eqref{eq:cRprop},
one can see that \eqref{eq:qu-string} reduces to 
\eqref{eq:classical-string} in the classical limit $\hbar\to0$. 
Note that the shift of $t_1$ in \eqref{eq:til-tk}
is important to recover the classical equation 
\eqref{eq:classical-string}.
This equation \eqref{eq:qu-string} determines the
$x$-dependence of $u(x)$. For instance, the string equation for the pure gravity
$t_0=-\cR_2$ is known as the Painlev\'{e} I equation. 
The so-called minimal string theory
(2d gravity coupled to a minimal model CFT) \cite{Seiberg:2004at,Seiberg:2003nm}
is obtained by turning on a finite number of couplings $t_k$,
in which case the string equation can be solved at least numerically 
\cite{Brezin:1990vd,Douglas:1990xv,Maldacena:2004sn}.

For the JT gravity case $t_n=\ga_n$ \eqref{eq:gammavalue1},
$P$ in \eqref{eq:P-diff} becomes
\begin{equation}
\begin{aligned}
 P=\Biggl[\hf\sin\bigl(2Q^\hf\bigr)\Biggr]_+.
\end{aligned} 
\label{eq:P-JT}
\end{equation}
In the classical limit,
this reduces to 
the spectral curve in \eqref{eq:curve} 
by the replacement $P\to y,Q\to\xi$.
\eqref{eq:P-JT} can be thought of as the ``quantum spectral curve'' for the JT gravity.
It would be interesting to study the property of \eqref{eq:P-JT} along the lines of 
\cite{Gukov:2011qp,Bouchard:2016obz}.

Let us consider the string equation \eqref{eq:qu-string}
for the JT gravity case.
We set $t_n=\ga_n$ for $n\geq1$ and leave $t_0$ as a free parameter.
Then \eqref{eq:qu-string} becomes
\begin{equation}
\begin{aligned}
 t_0=-\sum_{k=1}^\infty \frac{(-1)^k}{(k-1)!}\cR_k.
\end{aligned} 
\label{eq:string-JT}
\end{equation}
Due to the fact that infinitely many couplings $t_n$ are turned on,
\eqref{eq:string-JT} is no longer a differential equation for $u$;
it is rather thought to be a certain non-linear \textit{difference} equation for $u$.
Based on this expectation, 
we would like to write down the string equation \eqref{eq:string-JT}
in the form 
\begin{equation}
\begin{aligned}
 \hbar x=\sum_{n=1}^\infty \mathcal{D}_n(\del_{x_1},\cdots,\del_{x_n})
\prod_{i=1}^n u(x_i)\Bigr|_{x_i=x}.
\end{aligned} 
\end{equation}
The operator $\mathcal{D}_n$ can be found from the
recursion relation of $\cR_k$ \eqref{eq:rec_R}.
The first two terms are
\begin{equation}
\begin{aligned}
 \mathcal{D}_1&=\frac{\sin(\del_x)}{\del_x},\\
\mathcal{D}_2&=
\frac{\sin(\del_{x_1}+\del_{x_2})-\sin \del_{x_1}-\sin \del_{x_2}}{\del_{x_1}\del_{x_2}(\del_{x_1}+\del_{x_2})}.
\end{aligned} 
\end{equation}
Appearance of the exponentiated derivative in
$\sin(\del_x)=\frac{e^{\ri\del_x}-e^{-\ri\del_x}}{2\ri}$ indicates that $\mathcal{D}_n$
should be regarded as a \textit{difference} operator rather than a differential operator.
It would be interesting to find the general structure of $\mathcal{D}_n$.

%%%%%%%%%%%%%%%%%%%%%%%%%%%%%%%%%%%%%%%%%%%%%%%%%%%%%%%%%%%%%%%%%%%%%%%%
\section{Resolvent and wave functions}\label{app:BA}
%%%%%%%%%%%%%%%%%%%%%%%%%%%%%%%%%%%%%%%%%%%%%%%%%%%%%%%%%%%%%%%%%%%%%%%%

In this appendix we summarize useful properties of
the resolvent and the wave functions for the Schr\"odinger equation.

As discussed in \cite{BDY,BBT}, one can integrate
the equation for $R$ in \eqref{eq:R-eq} once. By multiplying $R$ to 
the first equation in \eqref{eq:R-eq} we find
\begin{equation}
\begin{aligned}
 0&=R\left[\qu R'''+(u-\xi)R'+\hf u' R\right]\\
&=\del_x \left[ \qu RR''-\frac{1}{8} R'^2+\hf (u-\xi)R^2\right].
\end{aligned} 
\end{equation}
This is integrated as 
\begin{equation}
\begin{aligned}
2RR''-R'^2+4(u-\xi)R^2=\text{const} .
\end{aligned} 
\label{eq:const}
\end{equation}
From the large $\xi$ behavior of $R$
\begin{equation}
\begin{aligned}
\lim_{\xi\to\infty} R= \xi^{-\hf}R_0=\hf \xi^{-\hf},
\end{aligned} 
\end{equation}
the constant on the right hand side of \eqref{eq:const}
is fixed to be $-1$
\begin{equation}
\begin{aligned}
 2RR''-R'^2+4(u-\xi)R^2= -1.
\end{aligned} 
\label{eq:R-eq2}
\end{equation}
From this equation, one can show that $\rt{R}$ satisfies
\begin{equation}
\begin{aligned}
 \bigl(Q-\xi\bigr)\rt{R}=-\frac{1}{4R^{3/2}}.
\end{aligned} 
\label{eq:R-half-eq}
\end{equation}

The resolvent $R$ can be written as a product of two functions
\begin{equation}
\begin{aligned}
 R=\psi_{+}\psi_{-},
\end{aligned} 
\end{equation}
where $\psi_\pm$ takes the form
\begin{equation}
\begin{aligned}
 \psi_\pm =\rt{R} e^{\pm S} .
\end{aligned} 
\end{equation}
Using the equation for $\rt{R}$ in \eqref{eq:R-half-eq}, one can show that
$\psi_\pm$ becomes a solution of the Schr\"{o}dinger equation $(Q-\xi)\psi_\pm =0$ provided that
$S$ satisfies \cite{BBT,Ambjorn:2005jx}
\begin{equation}
\begin{aligned}
 S'=\frac{1}{2R}.
\end{aligned} 
\end{equation}
This is integrated as
\begin{equation}
\begin{aligned}
 S(\xi,t_0)=\frac{1}{\hbar}\int^{t_0} dt_0 \frac{1}{2R[\xi,u(t_0)]}.
\end{aligned}
\label{eq:S-Rint} 
\end{equation}

Let us consider the classical limit of $S$.
On general ground, we expect that $S_{\text{cl}}$ is written as
\begin{equation}
\begin{aligned}
 S_{\text{cl}}=\frac{1}{\hbar}\int_0^\xi y(\xi')d\xi'
\end{aligned} 
\label{eq:Scl-yint}
\end{equation}
where $y$ is given by the classical limit of $P$ in \eqref{eq:P-diff}
\begin{equation}
\begin{aligned}
 y=-\sum_{k=1}^\infty \til{t}_k\frac{2^{k-1}}{(2k-1)!!}\xi^{k-1/2}.
\end{aligned} 
\end{equation}
Evaluating the integral in \eqref{eq:Scl-yint} we find
\begin{equation}
\begin{aligned}
S_{\text{cl}} &=-\frac{1}{\hbar}\sum_{k=1}^\infty \til{t}_k\frac{2^k\xi^{k+1/2}}{(2k+1)!!}.
\end{aligned} 
\label{eq:ydx-int}
\end{equation}
On the other hand, we can take the classical limit of \eqref{eq:S-Rint}
directly.
At the classical level $\hbar=0$, one can see from \eqref{eq:R-eq2}
that $R$ has a square-root branch cut
\begin{equation}
\begin{aligned}
 R_{\text{cl}}=\frac{1}{2\rt{\xi-u_0}}.
\end{aligned} 
\label{eq:Rcl}
\end{equation} 
Plugging \eqref{eq:Rcl} into \eqref{eq:S-Rint}
we find
\begin{equation}
\begin{aligned}
 S_{\text{cl}}&=\frac{1}{\hbar}\int dt_0 \rt{\xi-u_0}.
\end{aligned} 
\end{equation}
Using the classical string equation 
\begin{equation}
\begin{aligned}
 t_0=-\sum_{k=1}^\infty \til{t}_k\frac{u_0^k}{k!},
\end{aligned} 
\end{equation}
we can rewrite the $t_0$-integral to $u_0$-integral
\begin{equation}
\begin{aligned}
 S_{\text{cl}}
&=\frac{1}{\hbar}\int_0^{\xi} du_0 \frac{\del t_0}{\del u_0}\rt{\xi-u_0}\\
&=-\frac{1}{\hbar}\int_0^{\xi} du_0\sum_{k=1}^\infty \til{t}_k\frac{u_0^{k-1}}{(k-1)!}\rt{\xi-u_0}\\
&=-\frac{1}{\hbar}\sum_{k=1}^\infty \til{t}_k\frac{2^k\xi^{k+1/2}}{(2k+1)!!}.
\end{aligned} 
\end{equation}
This agrees with the integral of $yd\xi$ in \eqref{eq:ydx-int}.

For the JT gravity case $t_n=\ga_n$ we find
\begin{equation}
\begin{aligned}
 S_{\text{cl}}&=
-\frac{1}{\hbar}\sum_{k=1}^\infty \frac{(-1)^k}{(k-1)!}\frac{2^k\xi^{k+1/2}}{(2k+1)!!}
=\frac{1}{4\hbar}\Bigl[\sin(2\rt{\xi})-2\rt{\xi}\cos(2\rt{\xi})\Bigr],
\end{aligned} 
\label{eq:Scl-onshell}
\end{equation}
which reproduces the effective potential $V_{\text{eff}}(-\xi)$ 
in \eqref{eq:V-eff}, as expected.

%%%%%%%%%%%%%%%%%%%%%%%%%%%%%%%%%%%%%%%%%%%%%%%%%%%%%%%%%%%%%%%%%%%%%%%%


\begin{thebibliography}{99}
%%%%%%%%%%%%%%%%%%%%%%%%%%%%%%%%%%%%%%%%%%%%%%%%%%%%%%%%%%%%%%%%%%%%%%%%
\bibitem{Sachdev}
S.~Sachdev and J.~Ye,
``Gapless Spin-Fluid Ground State in a Random Quantum Heisenberg Magnet,''
\href{https://doi.org/10.1103/PhysRevLett.70.3339}{Phys.~Rev.~Lett. \textbf{70}, 3339 (1993)},
\href{https://arxiv.org/abs/cond-mat/9212030}{arXiv:cond-mat/9212030}.

\bibitem{kitaev}
A.~Kitaev,
``A simple model of quantum holography (part 1 and 2),''
Talks at KITP on
\href{http://online.kitp.ucsb.edu/online/entangled15/kitaev/}{April 7, 2015}
and
\href{http://online.kitp.ucsb.edu/online/entangled15/kitaev2/}{May 27, 2015}.

%\cite{Maldacena:2016hyu}
\bibitem{Maldacena:2016hyu} 
  J.~Maldacena and D.~Stanford,
  ``Remarks on the Sachdev-Ye-Kitaev model,''
\href{https://doi.org/10.1103/PhysRevD.94.106002}{Phys.\ Rev.\ D {\bf 94}, no. 10, 106002 (2016)},
\href{https://arxiv.org/abs/1604.07818}{[arXiv:1604.07818 [hep-th]]}.
  %%CITATION = doi:10.1103/PhysRevD.94.106002;%%

%\cite{Jackiw:1984je}
\bibitem{Jackiw:1984je} 
  R.~Jackiw,
  ``Lower Dimensional Gravity,''
\href{https://doi.org/10.1016/0550-3213(85)90448-1}{Nucl.\ Phys.\ B {\bf 252}, 343 (1985)}.
  %%CITATION = doi:10.1016/0550-3213(85)90448-1;%%

%\cite{Teitelboim:1983ux}
\bibitem{Teitelboim:1983ux} 
  C.~Teitelboim,
  ``Gravitation and Hamiltonian Structure in Two Space-Time Dimensions,''
\href{https://doi.org/10.1016/0370-2693(83)90012-6}{Phys.\ Lett.\  {\bf 126B}, 41 (1983)}.
  %%CITATION = doi:10.1016/0370-2693(83)90012-6;%%

%\cite{Almheiri:2014cka}
\bibitem{Almheiri:2014cka} 
  A.~Almheiri and J.~Polchinski,
  ``Models of AdS$_{2}$ backreaction and holography,''
\href{https://doi.org/10.1007/JHEP11(2015)014}{JHEP {\bf 1511}, 014 (2015)},
\href{https://arxiv.org/abs/1402.6334}{[arXiv:1402.6334 [hep-th]]}.
  %%CITATION = doi:10.1007/JHEP11(2015)014;%%

%\cite{Maldacena:2016upp}
\bibitem{Maldacena:2016upp} 
  J.~Maldacena, D.~Stanford and Z.~Yang,
  ``Conformal symmetry and its breaking in two dimensional Nearly Anti-de-Sitter space,''
\href{https://doi.org/10.1093/ptep/ptw124}{PTEP {\bf 2016}, no. 12, 12C104 (2016)},
\href{https://arxiv.org/abs/1606.01857}{[arXiv:1606.01857 [hep-th]]}.

%\cite{Jensen:2016pah}
\bibitem{Jensen:2016pah} 
  K.~Jensen,
  ``Chaos in AdS$_2$ Holography,''
\href{https://doi.org/10.1103/PhysRevLett.117.111601}{Phys.\ Rev.\ Lett.\  {\bf 117}, no. 11, 111601 (2016)},
\href{https://arxiv.org/abs/1605.06098}{[arXiv:1605.06098 [hep-th]]}.

%\cite{Engelsoy:2016xyb}
\bibitem{Engelsoy:2016xyb} 
  J.~Engels{\"o}y, T.~G.~Mertens and H.~Verlinde,
  ``An investigation of AdS$_{2}$ backreaction and holography,''
\href{https://doi.org/10.1007/JHEP07(2016)139}{JHEP {\bf 1607}, 139 (2016)},
\href{https://arxiv.org/abs/1606.03438}{[arXiv:1606.03438 [hep-th]]}.
  %%CITATION = doi:10.1007/JHEP07(2016)139;%%

%\cite{Saad:2019lba}
\bibitem{Saad:2019lba} 
  P.~Saad, S.~H.~Shenker and D.~Stanford,
  ``JT gravity as a matrix integral,''
\href{https://arxiv.org/abs/1903.11115}{arXiv:1903.11115 [hep-th]}.
  %%CITATION = ARXIV:1903.11115;%%

\bibitem{Mirzakhani}
M.~Mirzakhani,
``Simple geodesics and Weil-Petersson volumes of moduli spaces of bordered Riemann surfaces,''
\href{https://doi.org/10.1007/s00222-006-0013-2}{Invent.~Math. \textbf{167}, 179-222 (2007)}. 

%\cite{Eynard:2007fi}
\bibitem{Eynard:2007fi} 
  B.~Eynard and N.~Orantin,
  ``Weil-Petersson volume of moduli spaces, Mirzakhani's recursion and matrix models,''
\href{https://arxiv.org/abs/0705.3600}{arXiv:0705.3600 [math-ph]}.
  %%CITATION = ARXIV:0705.3600;%%

%\cite{Stanford:2017thb}
\bibitem{Stanford:2017thb} 
  D.~Stanford and E.~Witten,
  ``Fermionic Localization of the Schwarzian Theory,''
\href{https://doi.org/10.1007/JHEP10(2017)008}{JHEP {\bf 1710}, 008 (2017)},
\href{https://arxiv.org/abs/1703.04612}{[arXiv:1703.04612 [hep-th]]}.

%\cite{Eynard:2007kz}
\bibitem{Eynard:2007kz} 
  B.~Eynard and N.~Orantin,
  ``Invariants of algebraic curves and topological expansion,''
\href{https://doi.org/10.4310/CNTP.2007.v1.n2.a4}{Commun.\ Num.\ Theor.\ Phys.\  {\bf 1}, 347 (2007)},
\href{https://arxiv.org/abs/math-ph/0702045}{[math-ph/0702045]}.
  %%CITATION = doi:10.4310/CNTP.2007.v1.n2.a4;%%

%\cite{Gross:1989vs}
\bibitem{Gross:1989vs} 
  D.~J.~Gross and A.~A.~Migdal,
  ``Nonperturbative Two-Dimensional Quantum Gravity,''
\href{https://doi.org/10.1103/PhysRevLett.64.127}{Phys.\ Rev.\ Lett.\  {\bf 64}, 127 (1990)}.

%\cite{Gross:1989aw}
\bibitem{Gross:1989aw} 
  D.~J.~Gross and A.~A.~Migdal,
  ``A Nonperturbative Treatment of Two-dimensional Quantum Gravity,''
\href{https://doi.org/10.1016/0550-3213(90)90450-R}{Nucl.\ Phys.\ B {\bf 340}, 333 (1990)}.

%\cite{Douglas:1989ve}
\bibitem{Douglas:1989ve} 
  M.~R.~Douglas and S.~H.~Shenker,
  ``Strings in Less Than One-Dimension,''
\href{https://doi.org/10.1016/0550-3213(90)90522-F}{Nucl.\ Phys.\ B {\bf 335}, 635 (1990)}.

%\cite{Brezin:1990rb}
\bibitem{Brezin:1990rb} 
  E.~Brezin and V.~A.~Kazakov,
  ``Exactly Solvable Field Theories of Closed Strings,''
\href{https://doi.org/10.1016/0370-2693(90)90818-Q}{Phys.\ Lett.\ B {\bf 236}, 144 (1990)}.
  %%CITATION = doi:10.1016/0370-2693(90)90818-Q;%%

%\cite{Ginsparg:1993is}
\bibitem{Ginsparg:1993is} 
  P.~H.~Ginsparg and G.~W.~Moore,
  ``Lectures on 2-D gravity and 2-D string theory,''
\href{https://arxiv.org/abs/hep-th/9304011}{[hep-th/9304011]}.
  %%CITATION = HEP-TH/9304011;%%

%\cite{Seiberg:2004at}
\bibitem{Seiberg:2004at} 
  N.~Seiberg and D.~Shih,
  ``Minimal string theory,''
\href{https://doi.org/10.1016/j.crhy.2004.12.007}{Comptes Rendus Physique {\bf 6}, 165 (2005)},
\href{https://arxiv.org/abs/hep-th/0409306}{[hep-th/0409306]}.
  %%CITATION = doi:10.1016/j.crhy.2004.12.007;%%

%\cite{Seiberg:2003nm}
\bibitem{Seiberg:2003nm} 
  N.~Seiberg and D.~Shih,
  ``Branes, rings and matrix models in minimal (super)string theory,''
\href{https://doi.org/10.1088/1126-6708/2004/02/021}{JHEP {\bf 0402}, 021 (2004)},
\href{https://arxiv.org/abs/hep-th/0312170}{[hep-th/0312170]}.
  %%CITATION = doi:10.1088/1126-6708/2004/02/021;%%

%\cite{Banks:1989df}
\bibitem{Banks:1989df} 
  T.~Banks, M.~R.~Douglas, N.~Seiberg and S.~H.~Shenker,
``Microscopic and Macroscopic Loops in Nonperturbative Two-dimensional Gravity,''
\href{https://doi.org/10.1016/0370-2693(90)91736-U}{Phys.\ Lett.\ B {\bf 238}, 279 (1990)}.
  %%CITATION = doi:10.1016/0370-2693(90)91736-U;%%

\bibitem{zograf1}
P.~Zograf,
``On the large genus asymptotics of Weil-Petersson volumes,''
\href{https://arxiv.org/abs/0812.0544}{arXiv:0812.0544 [math.AG]}.

%\cite{Fateev:2000ik}
\bibitem{Fateev:2000ik} 
  V.~Fateev, A.~B.~Zamolodchikov and A.~B.~Zamolodchikov,
  ``Boundary Liouville field theory. 1. Boundary state and boundary two point function,''
\href{https://arxiv.org/abs/hep-th/0001012}{[hep-th/0001012]}.
  %%CITATION = HEP-TH/0001012;%%

%\cite{Teschner:2000md}
\bibitem{Teschner:2000md} 
  J.~Teschner,
  ``Remarks on Liouville theory with boundary,''
\href{https://doi.org/10.22323/1.006.0041}{PoS tmr {\bf 2000}, 041 (2000)},
\href{https://arxiv.org/abs/hep-th/0009138}{[hep-th/0009138]}.
  %%CITATION = doi:10.22323/1.006.0041;%%

%\cite{Dijkgraaf:2018vnm}
\bibitem{Dijkgraaf:2018vnm} 
  R.~Dijkgraaf and E.~Witten,
  ``Developments in Topological Gravity,''
\href{https://doi.org/10.1142/S0217751X18300296}{Int.\ J.\ Mod.\ Phys.\ A {\bf 33}, no. 30, 1830029 (2018)},
\href{https://arxiv.org/abs/1804.03275}{[arXiv:1804.03275 [hep-th]]}.
  %%CITATION = doi:10.1142/S0217751X18300296;%%

\bibitem{mulase}
M.~Mulase and B.~Safnuk,
``Mirzakhani's recursion relations, Virasoro constraints and the KdV hierarchy,''
\href{https://arxiv.org/abs/math/0601194}{arXiv:math/0601194 [math.QA]}.

\bibitem{Witten}
E.~Witten,
``Two-dimensional gravity and intersection theory on moduli space,''
\href{https://doi.org/10.4310/SDG.1990.v1.n1.a5}{Surveys Diff.~Geom.~\textbf{1}, 243-310 (1991)}.

\bibitem{Kontsevich}
M.~Kontsevich,
``Intersection theory on the moduli space of curves and the matrix Airy function,''
\href{https://doi.org/10.1007/BF02099526}{Comm.~Math.~Phys. \textbf{147}, 1-23 (1992)}.

%\cite{Itzykson:1992ya}
\bibitem{Itzykson:1992ya} 
  C.~Itzykson and J.~B.~Zuber,
  ``Combinatorics of the modular group. 2. The Kontsevich integrals,''
\href{https://doi.org/10.1142/S0217751X92002581}{Int.\ J.\ Mod.\ Phys.\ A {\bf 7}, 5661 (1992)},
\href{https://arxiv.org/abs/hep-th/9201001}{[hep-th/9201001]}.
  %%CITATION = doi:10.1142/S0217751X92002581;%%

\bibitem{Dikii}
I.~M.~Gelfand and L.~A.~Dikii,
``Asymptotic behavior of the resolvent of Sturm-Liouville equations and the algebra of the Korteweg-De Vries equations,''
\href{https://doi.org/10.1070/RM1975v030n05ABEH001522}{Russ.~Math.~Surveys {\bf 30}, 77-113, (1975)}.

\bibitem{DiFrancesco:1993cyw} 
  P.~Di Francesco, P.~H.~Ginsparg and J.~Zinn-Justin,
  ``2-D Gravity and random matrices,''  
\href{https://doi.org/10.1016/0370-1573(94)00084-G}{Phys.\ Rept.\  {\bf 254}, 1 (1995)},
\href{https://arxiv.org/abs/hep-th/9306153}{[hep-th/9306153]}.
  %%CITATION = doi:10.1016/0370-1573(94)00084-G;%%

%\cite{Makeenko:1991ec}
\bibitem{Makeenko:1991ec} 
  Y.~Makeenko and G.~W.~Semenoff,
  ``Properties of Hermitean matrix models in an external field,''
\href{https://doi.org/10.1142/S0217732391003985}{Mod.\ Phys.\ Lett.\ A {\bf 6}, 3455 (1991)}.
  %%CITATION = doi:10.1142/S0217732391003985;%%

%\cite{Ambjorn:1993sj}
\bibitem{Ambjorn:1993sj} 
  J.~Ambjorn and C.~F.~Kristjansen,
  ``From 1 matrix model to Kontsevich model,''
\href{https://doi.org/10.1142/S0217732393003263}{Mod.\ Phys.\ Lett.\ A {\bf 8}, 2875 (1993)},
\href{https://arxiv.org/abs/hep-th/9307063}{[hep-th/9307063]}.
  %%CITATION = doi:10.1142/S0217732393003263;%%

\bibitem{zograf-data}
P.~Zograf, unpublished.

\bibitem{BDY}
M.~Bertola, B.~Dubrovin and D.~Yang,
``Correlation functions of the KdV hierarchy and applications to intersection numbers over 
$\b{\mathcal{M}}_{g,n}$,''
\href{https://doi.org/10.1016/j.physd.2016.04.008}{Physica D \textbf{327}, 30-57 (2016)},
\href{https://arxiv.org/abs/1504.06452}{arXiv:1504.06452 [math-ph]}.

\bibitem{zograf2}
M.~Mirzakhani, P.~Zograf,
``Towards large genus asymtotics of intersection numbers on moduli spaces of curves,''
\href{https://arxiv.org/abs/1112.1151}{arXiv:1112.1151 [math.AG]}.

%\cite{Zamolodchikov:2001ah}
\bibitem{Zamolodchikov:2001ah} 
  A.~B.~Zamolodchikov and A.~B.~Zamolodchikov,
  ``Liouville field theory on a pseudosphere,''
\href{https://arxiv.org/abs/hep-th/0101152}{[hep-th/0101152]}.
  %%CITATION = HEP-TH/0101152;%%

%\cite{Garcia-Garcia:2016mno}
\bibitem{Garcia-Garcia:2016mno} 
  A.~M.~Garc\'{i}a-Garc\'{i}a and J.~J.~M.~Verbaarschot,
  ``Spectral and thermodynamic properties of the Sachdev-Ye-Kitaev model,''
\href{https://doi.org/10.1103/PhysRevD.94.126010}{Phys.\ Rev.\ D {\bf 94}, no. 12, 126010 (2016)},
\href{https://arxiv.org/abs/1610.03816}{[arXiv:1610.03816 [hep-th]]}.
  %%CITATION = doi:10.1103/PhysRevD.94.126010;%%

%\cite{Cotler:2016fpe}
\bibitem{Cotler:2016fpe} 
  J.~S.~Cotler G.~Gur-Ari, M.~Hanada, J.~Polchinski, P.~Saad, S.~H.~Shenker, 
D.~Stanford, A.~Streicher and M.~Tezuka,
  ``Black Holes and Random Matrices,''
  \href{https://doi.org/10.1007/JHEP05(2017)118}{JHEP {\bf 1705}, 118 (2017)}, 
\href{https://doi.org/10.1007/JHEP09(2018)002}{Erratum: [JHEP {\bf 1809}, 002 (2018)]},
\href{https://arxiv.org/abs/1611.04650}{[arXiv:1611.04650 [hep-th]]}.
  %%CITATION = doi:10.1007/JHEP09(2018)002, 10.1007/JHEP05(2017)118;%%

%\cite{Saad:2018bqo}
\bibitem{Saad:2018bqo} 
  P.~Saad, S.~H.~Shenker and D.~Stanford,
  ``A semiclassical ramp in SYK and in gravity,''
\href{https://arxiv.org/abs/1806.06840}{arXiv:1806.06840 [hep-th]}.
  %%CITATION = ARXIV:1806.06840;%%

\bibitem{saad}
P.~Saad,
``Late Time Correlation Functions, Baby Universes, and ETH in JT Gravity,''
\href{https://arxiv.org/abs/1910.10311}{arXiv:1910.10311 [hep-th]}.

%\cite{Maldacena:2004sn}
\bibitem{Maldacena:2004sn} 
  J.~M.~Maldacena, G.~W.~Moore, N.~Seiberg and D.~Shih,
  ``Exact vs. semiclassical target space of the minimal string,''
\href{https://doi.org/10.1088/1126-6708/2004/10/020}{JHEP {\bf 0410}, 020 (2004)},
\href{https://arxiv.org/abs/hep-th/0408039}{[hep-th/0408039]}.
  %%CITATION = doi:10.1088/1126-6708/2004/10/020;%%

%\cite{Gaiotto:2003yb}
\bibitem{Gaiotto:2003yb} 
  D.~Gaiotto and L.~Rastelli,
  ``A Paradigm of open / closed duality: Liouville D-branes and the Kontsevich model,''
\href{https://doi.org/10.1088/1126-6708/2005/07/053}{JHEP {\bf 0507}, 053 (2005)},
\href{https://arxiv.org/abs/hep-th/0312196}{[hep-th/0312196]}.

\bibitem{Bertola} 
M.~Bertola and M.~Cafasso,
``Universality of the matrix Airy and Bessel functions at spectral edges of unitary ensembles,''
\href{https://doi.org/10.1142/S2010326317500101}{Random Matrices: Theory Appl. {\bf 06}, 
1750010 (2017)},
\href{https://arxiv.org/abs/1610.06108}{arXiv:1610.06108 [math-ph]}.

%\cite{Garcia-Garcia:2017pzl}
\bibitem{Garcia-Garcia:2017pzl} 
  A.~M.~Garc\'{i}a-Garc\'{i}a and J.~J.~M.~Verbaarschot,
  ``Analytical Spectral Density of the Sachdev-Ye-Kitaev Model at finite N,''
\href{https://doi.org/10.1103/PhysRevD.96.066012}{Phys.\ Rev.\ D {\bf 96}, no. 6, 066012 (2017)},
\href{https://arxiv.org/abs/1701.06593}{[arXiv:1701.06593 [hep-th]]}.
  %%CITATION = doi:10.1103/PhysRevD.96.066012;%%

%\cite{Berkooz:2018jqr}
\bibitem{Berkooz:2018jqr} 
  M.~Berkooz, M.~Isachenkov, V.~Narovlansky and G.~Torrents,
  ``Towards a full solution of the large N double-scaled SYK model,''
\href{https://doi.org/10.1007/JHEP03(2019)079}{JHEP {\bf 1903}, 079 (2019)},
\href{https://arxiv.org/abs/1811.02584}{[arXiv:1811.02584 [hep-th]]}.
  %%CITATION = doi:10.1007/JHEP03(2019)079;%%

\bibitem{Iliev}
P.~Iliev,
``On the heat kernel and the Korteweg--de Vries hierarchy ,''
\href{https://doi.org/10.5802/aif.2154}{Annales de l'Institut Fourier 
{\bf55}, 2117-2127 (2005)}. 

\bibitem{OkuyamaSakai}
K.~Okuyama and K.~Sakai, in preparation.

%\cite{Stanford:2019vob}
\bibitem{Stanford:2019vob} 
  D.~Stanford and E.~Witten,
  ``JT Gravity and the Ensembles of Random Matrix Theory,''
\href{https://arxiv.org/abs/1907.03363}{arXiv:1907.03363 [hep-th]}.
  %%CITATION = ARXIV:1907.03363;%%

%\cite{Iliesiu:2019lfc}
\bibitem{Iliesiu:2019lfc} 
  L.~V.~Iliesiu,
  ``On 2D gauge theories in Jackiw-Teitelboim gravity,''
\href{https://arxiv.org/abs/1909.05253}{arXiv:1909.05253 [hep-th]}.

%\cite{Maldacena:2019cbz}
\bibitem{Maldacena:2019cbz} 
  J.~Maldacena, G.~J.~Turiaci and Z.~Yang,
  ``Two dimensional Nearly de Sitter gravity,''
\href{https://arxiv.org/abs/1904.01911}{arXiv:1904.01911 [hep-th]}.
  %%CITATION = ARXIV:1904.01911;%%

%\cite{Cotler:2019nbi}
\bibitem{Cotler:2019nbi} 
  J.~Cotler, K.~Jensen and A.~Maloney,
  ``Low-dimensional de Sitter quantum gravity,''
\href{https://arxiv.org/abs/1905.03780}{arXiv:1905.03780 [hep-th]}.
  %%CITATION = ARXIV:1905.03780;%%

\bibitem{Okounkov}
A.~Okounkov,
``Generating functions for intersection numbers on moduli spaces of curves,''
\href{https://doi.org/10.1155/S1073792802110099}
{International Mathematics Research Notices {\bf 18} (2002)},
\href{https://arxiv.org/abs/math/0101201}{arXiv:math/0101201 [math.AG]}.

%\cite{Maldacena:2018lmt}
\bibitem{Maldacena:2018lmt} 
  J.~Maldacena and X.~L.~Qi,
  ``Eternal traversable wormhole,''
\href{https://arxiv.org/abs/1804.00491}{arXiv:1804.00491 [hep-th]}.
  %%CITATION = ARXIV:1804.00491;%%

%\cite{Okuyama:2019xvg}
\bibitem{Okuyama:2019xvg} 
  K.~Okuyama,
  ``Replica symmetry breaking in random matrix model: a toy model of wormhole networks,''
\href{https://arxiv.org/abs/1903.11776}{arXiv:1903.11776 [hep-th]}.
  %%CITATION = ARXIV:1903.11776;%%

%\cite{Okuyama:2018yep}
\bibitem{Okuyama:2018yep} 
  K.~Okuyama,
  ``Spectral form factor and semi-circle law in the time direction,''
\href{https://doi.org/10.1007/JHEP02(2019)161}{JHEP {\bf 1902}, 161 (2019)},
\href{https://arxiv.org/abs/1811.09988}{[arXiv:1811.09988 [hep-th]]}.
  %%CITATION = doi:10.1007/JHEP02(2019)161;%%

%\cite{Douglas:1989dd}
\bibitem{Douglas:1989dd} 
  M.~R.~Douglas,
  ``Strings in Less Than One-dimension and the Generalized KdV Hierarchies,''
\href{https://doi.org/10.1016/0370-2693(90)91716-O}{Phys.\ Lett.\ B {\bf 238}, 176 (1990)}.
  %%CITATION = doi:10.1016/0370-2693(90)91716-O;%%

%\cite{Brezin:1990vd}
\bibitem{Brezin:1990vd} 
  E.~Brezin, E.~Marinari and G.~Parisi,
  ``A Nonperturbative ambiguity free solution of a string model,''
\href{https://doi.org/10.1016/0370-2693(90)91590-8}{Phys.\ Lett.\ B {\bf 242}, 35 (1990)}.
  %%CITATION = doi:10.1016/0370-2693(90)91590-8;%%

%\cite{Douglas:1990xv}
\bibitem{Douglas:1990xv} 
  M.~R.~Douglas, N.~Seiberg and S.~H.~Shenker,
  ``Flow and Instability in Quantum Gravity,''
\href{https://doi.org/10.1016/0370-2693(90)90333-2}{Phys.\ Lett.\ B {\bf 244}, 381 (1990)}.
  %%CITATION = doi:10.1016/0370-2693(90)90333-2;%%

%\cite{Gukov:2011qp}
\bibitem{Gukov:2011qp} 
  S.~Gukov and P.~Sulkowski,
  ``A-polynomial, B-model, and Quantization,''
\href{https://doi.org/10.1007/JHEP02(2012)070}{JHEP {\bf 1202}, 070 (2012)},
\href{https://arxiv.org/abs/1108.0002}{[arXiv:1108.0002 [hep-th]]}.
  %%CITATION = doi:10.1007/JHEP02(2012)070;%%

%\cite{Bouchard:2016obz}
\bibitem{Bouchard:2016obz} 
  V.~Bouchard and B.~Eynard,
  ``Reconstructing WKB from topological recursion,''
\href{https://doi.org/10.5802/jep.58}{Journal de l'Ecole polytechnique -- Mathematiques, 4 (2017), p.
  845-908},
\href{https://arxiv.org/abs/1606.04498}{[arXiv:1606.04498 [math-ph]]}.
  %%CITATION = doi:10.5802/jep.58;%%

\bibitem{BBT}
O.~Babelon, D.~Bernard and M.~Talon,
``Introduction to Classical Integrable Systems,''
Cambridge University Press (2007).

%\cite{Ambjorn:2005jx}
\bibitem{Ambjorn:2005jx} 
  J.~Ambjorn and R.~A.~Janik,
  ``The Emergence of noncommutative target space in noncritical string theory,''
\href{https://doi.org/10.1088/1126-6708/2005/08/057}{JHEP {\bf 0508}, 057 (2005)},
\href{https://arxiv.org/abs/hep-th/0506197}{[hep-th/0506197]}.
  %%CITATION = doi:10.1088/1126-6708/2005/08/057;%%

%last

\end{thebibliography}
\end{document}